\documentclass[11pt]{article}
\usepackage[utf8]{inputenc}
\usepackage[margin=1in]{geometry}

\usepackage[T1]{fontenc}
\usepackage{lmodern}

\usepackage{microtype}
\usepackage{graphicx}
\usepackage{subfigure}
\usepackage{booktabs}
\usepackage{natbib}
\setcitestyle{open={(},close={)}}

\usepackage{setspace}

\usepackage{amsmath}
\usepackage{amssymb,amsmath,amsthm}

\usepackage{mathtools}
\usepackage{amsthm}
\usepackage{algorithmic}
\usepackage{graphicx}
\usepackage{textcomp}
\usepackage{xcolor}
\usepackage{amsfonts}
\usepackage{epsfig}
\usepackage{algorithm}
\usepackage{wrapfig}
\usepackage{balance}
\usepackage{url}
\usepackage{mathtools}
\usepackage{natbib}
\usepackage{multirow}
\usepackage{soul,comment}

\usepackage[textsize=tiny]{todonotes}

\usepackage[colorlinks=true,citecolor=blue]{hyperref}

\usepackage{amsmath,amsthm,amsfonts,amssymb,mathdots,array,mathrsfs,bm,bbm,stmaryrd,graphicx,subfigure,xcolor}

\usepackage{breakcites}

\usepackage[T1]{fontenc}
\usepackage{enumerate}
\usepackage{inputenc}

\usepackage{graphicx} 
\usepackage{subfigure}

\usepackage{booktabs,balance}
\usepackage{rotating}
\usepackage{boldline}
\usepackage{makecell}
\usepackage{multirow}
\usepackage{balance}

\usepackage{tikz}

\newcommand{\bsmat}{\begin{bmatrix} }
\newcommand{\esmat}{\end{bmatrix} }

\usepackage{environ}
\NewEnviron{smallequation}{%
    \begin{equation}
    \scalebox{0.97}{$\BODY$}
    \end{equation}
    }
    \NewEnviron{smallalign}{%
    \begin{equation}
    \scalebox{0.97}{$\BODY$}
    \end{equation}
    }

\usepackage{multirow}
\usepackage{caption}
\usepackage{enumitem}
\usepackage{soul}
\usepackage{wrapfig,lipsum}

\usepackage{hyperref}
\usepackage{array}
\newcolumntype{C}[1]{>{\centering\let\newline\\\arraybackslash\hspace{0pt}}m{#1}}
\begin{document}

\title{\bf\huge Asymmetric Hashing for Fast Ranking via Neural Network Measures}

\author{\vspace{0.1in}\\\textbf{Khoa Doan, Shulong Tan, Weijie Zhao,  Ping Li} \\\\
Cognitive Computing Lab\\
Baidu Research\\
10900 NE 8th St. Bellevue, WA 98004, USA\\\\
  \texttt{\{khoadoan106,  tanshulong2011, zhaoweijie12, pingli98\}@gmail.com}
}

\date{}
\maketitle

\begin{abstract}\vspace{0.3in}
\noindent Fast item ranking is an important task in recommender systems. In previous works, graph-based Approximate Nearest Neighbor (ANN) approaches have demonstrated good performance on item ranking tasks with generic searching/matching measures (including complex measures such as neural network measures). However, since these ANN approaches must go through the neural measures several times during ranking, the computation is not practical if the neural measure is a large network.  On the other hand, fast item ranking using existing hashing-based approaches, such as Locality Sensitive Hashing (LSH), only works with a limited set of measures, such as cosine and Euclidean distance. Previous learning-to-hash approaches are also not suitable to solve the fast item ranking problem since they can take a significant amount of time and computation to train the hash functions due to a large number of possible training pairs in this problem. Hashing approaches, however, are attractive because they provide a principle and efficient way to retrieve candidate items. In this paper, we propose a simple and effective  learning-to-hash approach for the fast item ranking problem that can be used for any type of measure, including neural network measures. Specifically, we solve this problem with an asymmetric hashing framework based on discrete inner product fitting. We learn a pair of related hash functions that map heterogeneous objects (e.g., users and items) into a common discrete space where the inner product of their binary codes reveals their true similarity defined via the original searching measure. The fast ranking problem is reduced to an ANN search via this asymmetric hashing scheme. Then, we propose a sampling strategy to efficiently select relevant and contrastive samples to train the hashing model. We empirically validate the proposed method against the existing state-of-the-art fast item ranking methods in several combinations of non-linear searching functions and prominent datasets.
\end{abstract}

\newpage

\section{Introduction} \label{sec:intro}

The rapid growth of digital data brings plenty of opportunities but also presents many challenges, especially in recommendation, advertising, and information retrieval systems. In these systems, deep neural networks have been shown to effectively capture the complex, semantic relationships between heterogeneous objects~\citep{severyn2015learning,guo2016deep,dehghani2017neural,xiong2017end,tay2018latent,fan2019mobius,fei2021gemnn}. To obtain better model capacities, complex similarity/matching functions--say in the forms of neural networks--are usually exploited, instead of simple functions, such as $\ell_2$-distance, cosine similarity, or inner product. However, online ranking using these advanced, non-linear similarity functions is very challenging~\citep{covington2016deep,chen2017reading,mitra2018introduction,chang2020pre,tan2020fast} since it is too time-consuming to rank all data in real-time via these complex measures.

Efficient, sub-linear ranking via neural network-based measures is called \textbf{fast neural ranking}~\citep{mitra2018introduction,tan2020fast}. Sub-linear time neural ranking is considered challenging because the ranking measures (in the forms of neural networks) are too complicated for traditional fast vector ranking methods, i.e., approximate nearest neighbor (ANN) search algorithms. ANN algorithms include hashing-based methods, graph-based methods, tree-based methods, quantization-based methods, etc.~\citep{friedman1975algorithm,friedman1977algorithm, indyk1998approximate,broder1998min, gionis1999similarity,charikar2002similarity, jegou2011product, shrivastava2012fast, datar2004locality, malkov2020efficient,zhao2020song}. Traditional ANN search methods usually focus on some specific and simple ranking/searching measures. For example, data-independent hashing methods are commonly  used for ANN search with respect to the cosine similarity~\citep{ goemans1995improved,charikar2002similarity},  the chi-square similarity~\citep{li2013sign}, the (binary) Jaccard similarity~\citep{broder1998min,li2022c}, and the general (real-valued) Jaccard similarity~\citep{manasse2010consistent,ioffe2010improved,li2017linearized}. Recently, sub-linear ranking under inner product becomes popular in both academic and industry, due to its wide applications in recommendation and ads ranking. This problem is referred to as the Maximum Inner Product Search (MIPS)~\citep{ram2012maximum,bachrach2014speeding,shrivastava2014asymmetric,shrivastava2015asymmetric,morozov2018non,zhou2019mobius}.  Similar to traditional ANN algorithms for fast vector search,  the MIPS methods are specially  designed for the MIPS problem and are not straightforward to extend to fast neural ranking.

\vspace{0.1in}

To rank via neural network-based measures, \cite{tan2020fast} extend the definition of the traditional ANN search to a generic setting, called \textbf{Optimal Binary Function Search (OBFS)}:
Let $X$ and $Y$ be subsets of Euclidean spaces, given a data set $S = \{x_1, \dots, x_n\} \subset X$ and a continuous \textbf{binary function} (i.e., the similarity or matching function), $f: X\times Y \to \mathbb R$, given $q\in Y$, OBFS aims to find:
    \begin{align} \label{eq:obfs}
    \arg\max_{x_i\in S} f(x_i, q).
    \end{align}
In this definition, they consider an arbitrary binary function as the ranking measure, no matter it is linear (e.g., $\ell_2$ or cosine) or non-linear (e.g., a neural network). A neural network-based measure can be considered as a special case of the binary function and the fast neural ranking problem is a sub-problem of OBFS. Because existing ANN search methods are designed for specific metric measures, they are not suitable to solve the OBFS task. Furthermore, extending the existing ANN solutions to OBFS can be challenging because they generally do not have access to the original pairwise dataset, denoted as $\mathcal{D}_{orig}$, that is used to train $f$; for example, $f$ may be trained and provided by a third-party provider as an Off-the-shelf product. It is impossible to re-train the ranking model.

Besides the definition, \citet{tan2020fast} also provides a solution for fast OBFS, called Search on L2 Graph (SL2G). They extend traditional graph-based fast vector searching algorithms by constructing graph indices in the $\ell_2$-distance but searching or ranking according to the binary function $f$.
SL2G is the first approximate solution for sub-linear neural ranking or the generic OBFS problem. In SL2G, each given query is only compared with a subset of base data instead of the whole dataset, which reduces the time complexity significantly. Although SL2G is a sub-linear method, it is still required to go through the ranking measure hundreds of times to promise high retrieval recalls. For large and/or deep neural network measures (e.g., a BERT-style model~\citep{devlin2019bert,chang2020pre}), such repeated computation is not practical~for~online~services.

This paper solves the fast neural ranking problem by learning two related hash functions, one for the query and one for the item, whose inner product approximates the original ranking measure $f$. Given a query, the item ranking via $f$ is replaced with a fast ranking using the two hash functions.
Hash ranking is efficient as the distance computation is very fast, using only an XOR and a bit count computations, each of which typically requires one CPU cycle, in contrast to the expensive computation using the neural network measure $f$ of SL2G. In addition, ranking over the searching data can be performed in memory as the hash codes are memory-efficient (e.g., a million of 32-bit hash codes only occupy 4MB of memory).

Existing  works in the hashing domain are typically ineffective to solve the OBFS problem. Single-modal hashing algorithms are designed to deal with specific measures~\citep{broder1998min,indyk1998approximate,gionis1999similarity,charikar2002similarity,salakhutdinov2009semantic,manasse2010consistent,ioffe2010improved,shrivastava2012fast,gong2013iterative,li2013sign,datar2004locality,shen2015supervised,lin2016learning,huang2016unsupervised,huang2017unsupervised,li2017linearized,cao2018hashgan,yang2018supervised,li2022c}, instead of arbitrary measures (i.e., $f$) such as neural networks. Other cross-modal hashing algorithms~\citep{jiang2017deep,li2018self,zhang2018attention,gu2019adversary,kang2019candidate} learn the hash functions as part of the training process of the binary function $f$. They are not suitable to solve the OBFS problem because, in OBFS, we cannot assume how $f$ is trained. While it is possible to enumerate a pairwise dataset, denoted as $\mathcal{D}_{app}$, similar to $\mathcal{D}_{orig}$, from a set of queries $q \in Y$ and a set of items $x \in S$ that are collected in the online applications and train the hash functions using these algorithms, such enumerations result in a large number of training instances. For example, one million objects in each domain produce one trillion pairwise inputs. Training hash functions on such a large number of inputs using the aforementioned methods can lead to poor convergence of the learning process and take an unrealistic amount of time and computation. Note that $\mathcal{D}_{orig}$ and $\mathcal{D}_{app}$ are assumed to come from the same data distribution.

\begin{table*}[b!]
\centering
\caption{Characteristics of potential solutions to the OBFS problem. $\times$ or \checkmark~indicates the method lacks or has the specific characteristic, respectively. Our  FLORA   satisfies all the characteristics by employing an asymmetric hashing scheme and a novel training procedure. We compare FLORA with traditional ANN,  single-modal hashing, cross-modal hashing, and SL2G~\citep{tan2020fast}.\vspace{-0.1in}}
\begin{tabular}{|l|C{2.3cm}|C{2.3cm}|C{2.3cm}|C{1.0cm}|C{1.4cm}|}
\hline
       & \begin{tabular}[t]{@{}c@{}}Classic ANN\end{tabular} & \begin{tabular}[t]{@{}c@{}}Single-modal\\Hashing\end{tabular} &
       \begin{tabular}[t]{@{}c@{}}Cross-modal\\Hashing\end{tabular} & \begin{tabular}[t]{@{}c@{}}SL2G\end{tabular} & \begin{tabular}[t]{@{}c@{}}FLORA \end{tabular} \\ \hline
\textbf{Works on arbitrary $f$}  & $\times$   & $\times$   & \checkmark  & \checkmark  & \checkmark    \\ \hline
\textbf{Not dependent on $\mathcal{D}_{orig}$}  & \checkmark   &  \checkmark   &  $\times$ & \checkmark  & \checkmark    \\
\hline
\textbf{Ranking without $f$}  &  \checkmark    & \checkmark   & \checkmark  &  $\times$  & \checkmark    \\
\hline
\end{tabular}
\label{tbl:obfs_potential_solutions}\vspace{-0.1in}
\end{table*}

To address the above limitations of the existing methods for the fast neural ranking problem, in this paper, we propose \textbf{FLORA}, a \textbf{F}ast \textbf{L}earning t\textbf{O} \textbf{RA}nk method. The advantages of FLORA over the potential solutions for the OBFS  problem are shown in Table~\ref{tbl:obfs_potential_solutions}. FLORA's hashing model  learns two related hash functions to map objects from two heterogeneous domains to discrete representations whose inner product captures the objects' similarity defined by the binary function $f$.
Specifically, we reduce the ranking problem via the original binary function $f$ to an approximate nearest neighbor search via an asymmetric pair of hash functions. The model is asymmetric in the sense that one hash function is used to preprocess the data in the database (e.g., items), while the other hash function is used to transform the query (e.g., a user).
This framework can approximate any form of the binary function.
To train the hash functions, we propose a novel sampling strategy that efficiently picks input pairs out of the large number of all possible pairs ($n_1 \times n_2$) so that the model training converges faster to a good performance.

\vspace{0.1in}

In summary, the contributions of this paper are as below:
\begin{itemize}
    \item Propose a hashing neural network that captures the relationship of the heterogeneous objects through a pair of asymmetric hash functions.
    The proposed model is learned to preserve the semantic similarity of the inputs defined by the binary function and to generate balanced hash codes via three novel objective functions.
    \item Propose a novel strategy to efficiently and effectively sample data pairs to accelerate the training process and find better local optima of the model.
    \item Demonstrate the effectiveness of FLORA on various complex neural network-based matching functions and several benchmark datasets used in the recommendation domain. FLORA produces rankings with higher recalls and is significantly better in both computation and space, compared to existing methods.
	\end{itemize}

\vspace{0.1in}

\noindent \textbf{Paper  organization.} \ In Section~\ref{sec:relwork}, we review the fast neural ranking problem and its potential solutions.
Then we introduce the proposed hashing method, FLORA, for the fast neural ranking problem. In Section~\ref{sec:exp}, we study three complex ranking measures and conduct experiments. The entire paper is concluded in Section~\ref{sec:conclusion}.

\section{Related Work}
\label{sec:relwork}
In this section, we will first introduce some classic neural network-based ranking models and their applications. Then, the fast neural ranking problem will be reviewed in the context of the general top-k search. Finally, we will discuss some potential solutions for fast neural ranking and explain their limitations, then motivate the proposed methodology.

\subsection{Neural Network-based Ranking Models}
\label{sec:nn_measures}
Recently,  neural network-based models have been proposed for the ranking problems such as recommendation, information retrieval, ads ranking, and retrieval-based question answering~\citep{xiong2017end,  chen2017reading,devlin2019bert,fan2019mobius, huang2019knowledge}. Neural network-based representation learning models~\citep{huang2013learning, shen2014learning, xue2017deep,zamani2018neural,sun2018logician} show superior performance in feature learning. Heterogeneous objects can be embedded from their raw feature spaces into a common latent space. In this space, different objects (e.g., users and items, questions and answers) can be compared by some classic measures, such as cosine, $\ell_2$ distance, or inner product.

\newpage

To achieve even greater capacities in modeling complicated relationships between heterogeneous objects, previous works try to learn meaningful matching functions (so-called \textbf{matching function learning}), especially in forms of deep neural networks~\citep{lu2013deep,severyn2015learning,dehghani2017neural,xiong2017end,tay2018latent}, instead of the classic measures. These works assume that there exists a mapping that can capture the complicated relationship (e.g., non-linear and non-convex) between heterogeneous objects. Therefore, the goal of matching function learning is to learn the mapping by well-designed models. For the matching models (i.e., matching functions to be learned), Multi-Layer Perceptrons (MLPs) are commonly adopted~\citep{severyn2015learning,dehghani2017neural,tay2018latent,yu2022s2mlp}. Recently, BERT-style models are widely used for language modeling~\citep{devlin2019bert,chang2020pre}. It has been shown that matching function learning methods outperform simple representation learning methods in complicated ranking tasks~\citep{xu2018deep}.

\subsection{Fast Neural Ranking}

For online ranking services, search efficiency is as important as search effectiveness. Fast search algorithms construct some forms of index structures in advance in order to accelerate the online search, such as the inverted index for free text and LSH tables for vector data. Traditional fast search problems restrict the searching measures to metric measures (e.g., $\ell_2$ distance) or some simple non-metric measures, such as inner product (which is widely exploited in recommender systems).

OBFS (i.e., Eq.~\eqref{eq:obfs})~\citep{tan2020fast} is a generic definition for fast top-$K$ search. Theoretically, an OBFS task can choose any binary function as the search measure, whether it is linear or non-linear, metric or non-metric, convex or non-convex, or symmetric or asymmetric. Traditional ANN search and MIPS are special cases of OBFS. Recently, neural network-based ranking measures~\citep{guo2016deep, lu2013deep,dehghani2017neural,chang2020pre} increasingly become popular examples of searching binary functions, as introduced in Section \ref{sec:nn_measures}. These kinds of neural network-based searching functions are usually non-convex. In this paper, we focus on the generic, fast OBFS problem, especially the so-called fast neural ranking problem of how to speed up the searching process under neural network-based measures for online ranking services.




The Search on L2 Graph (SL2G)~\citep{tan2020fast} is designed for the generic fast OBFS problem. The basic idea of SL2G is: no matter what the given binary function $f$ is, they construct a Delaunay graph (or an approximate one) with respect to the $\ell_2$ distance in the indexing step; the  $\ell_2$ distance is defined on searching/base data $X$ and is, therefore, independent of the queries. Then SL2G performs the greedy search on this graph via the binary function $f$ at query time. To achieve higher recalls, SL2G must invoke the ranking function hundreds of times. For such reason, SL2G is not practical when the binary functions are complex networks.

\subsection{Learning to Hash}
Existing supervised hashing approaches~\citep{shen2015supervised,yang2018supervised,cao2018hashgan,gui2018r2sdh, jiang2018asymmetric,doan2022one,doan2020image} are designed to deal with ANN search.
Most of these methods assume that query and database objects are homogeneous  (e.g., both are images), but the matching function may have heterogeneous pair of inputs (e.g., user and item). While cross-modal hashing approaches~\citep{jiang2017deep,li2018self,zhang2018attention,gu2019adversary,kang2019candidate} learn the hash functions for heterogeneous objects, they are designed to use pairwise annotated training data. To employ these methods for the OBFS problem, we need to enumerate the pairwise data using the binary function $f$ and the separate sets of queries (i.e., users) and database (i.e., items) since we do not have access to the original pairwise training data that is used to train $f$. Such enumerations can result in a large number of training instances ($n_1 \times n_2$; e.g., 1 million users and 1 million items results in one trillion training instances), which makes it impractical to train these models.


\section{Fast Neural Ranking with Hashing}
\label{sec:proposed}

We solve the fast neural ranking problem by learning low-dimensional discrete hash codes whose distances approximate the original similarity measure defined by the continuous binary function $f$. The basic idea of FLORA is: 1) no matter what binary function $f$ is, we can always learn a pair of related hash functions $H_1$ and $H_2$ that together approximate the similarity defined by $f$ and 2) exact searching using $f$ is reduced to approximate searching in the discrete space using the distance between the discrete outputs of $H_1$ and $H_2$; this searching approach avoids the possibly complex and expensive computation of the original binary function $f$.

Next, we describe how the hash functions $H_1$ and $H_2$ are learned, and how the trained hash functions are used for fast neural~ranking.

\subsection{Hash Function Learning}

\begin{figure}[b!]
 \vspace{-0.1in}
    \centering
    \includegraphics[width=3.3in]{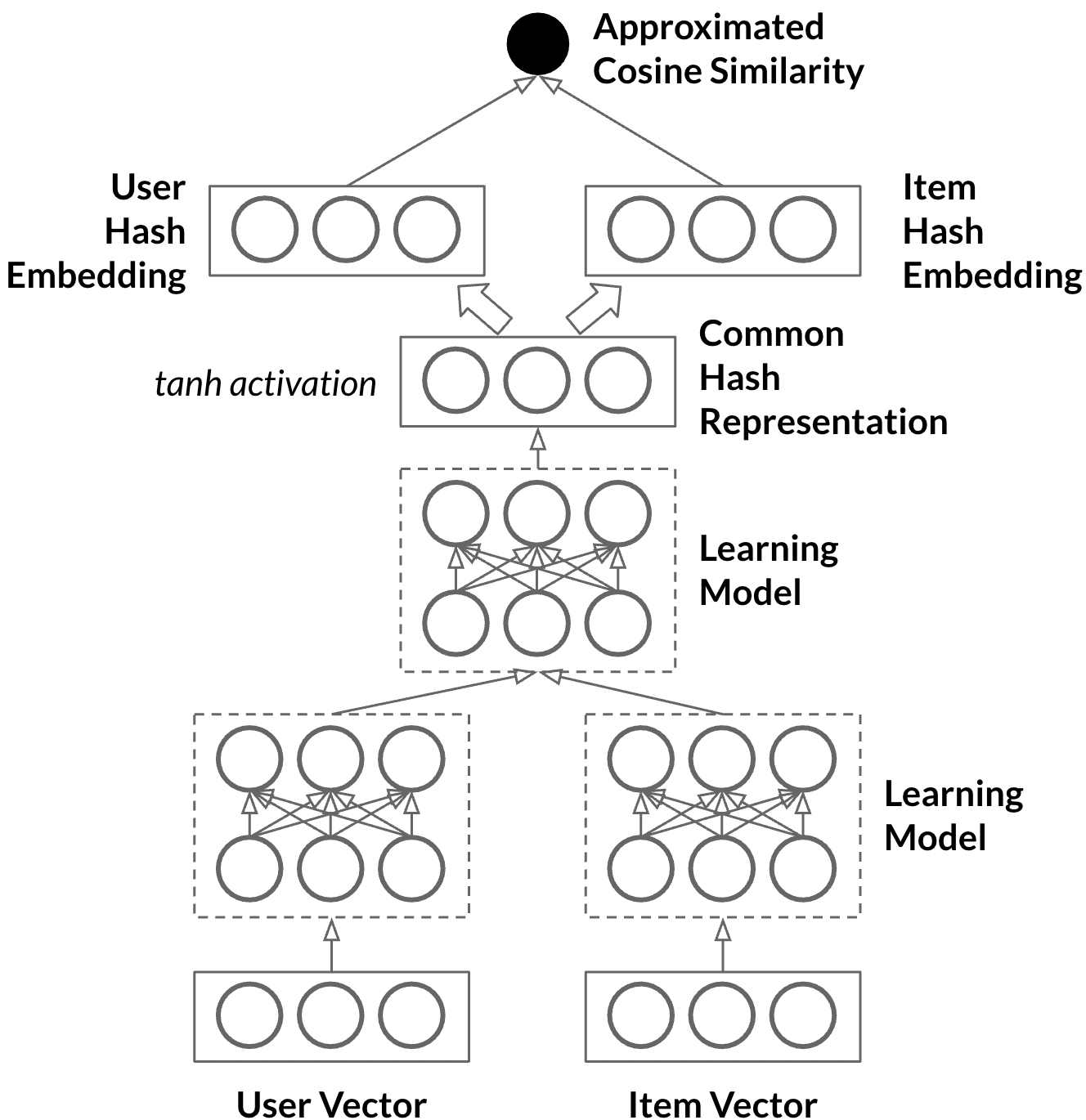}

    \vspace{-0.1in}

    \caption{FLORA learning-to-hash network. Item vectors and user vectors are the inputs. These vectors are different from raw features in original pairwise dataset, $\mathcal{D}_{orig}$.}
    \label{fig:fltr_network}
    \vspace{-0.1in}
\end{figure}

The input of the binary function $f$ consists of two domains (for example, the user and item domains in recommendation applications). In OBFS, we are given a trained binary function $f$ that computes the similarity between~a~user~and~an~item.

To approximate $f$, FLORA represents each user $u$ and item $v$ using two separate hash functions, denoted by  $H_1: u \rightarrow \{-1,1\}^m$ and $H_2: v \rightarrow \{-1,1\}^m$, respectively,
where $m$ is the dimension of the discrete spaces (i.e., the number of bits). To model the relationship between $H_1$ and $H_2$, we extend the two-tower neural network by incorporating a set of shared embedding layers, as presented in Figure~\ref{fig:fltr_network}. Each user input and item input first go into two separate MLP networks, denoted as $g_1$ and $g_2$, respectively, to extract the appropriate features for each domain. Then, the extracted features are embedded into the same discrete space through a shared MLP network, denoted as $g_0$. Equivalently, $H_1=g_0 \circ g_1$ and $H_2=g_0 \circ g_2$. In this space, the relationship between two objects is defined via the cosine similarity, which is also linearly related to their Hamming distance. This model can learn to extract different representations of the two domains while finally capturing high-order interactions between the two objects in the common discrete space.


The objective function of FLORA consists of three parts. The first part, denoted as $\mathcal{L}_c$, is the consistency loss that penalizes the difference between the predicted similarity score computed in the discrete space and the target similarity score defined by the binary function $f$. The consistency loss is defined as follows:
\begin{equation}
   \mathcal{L}_c  = E_{u, v} ||f(v, u) - cosine(H_1(u), H_2(v))||_2
    \label{eqn:consistency_objective}
\end{equation}
{where $cosine(H_1(u), H_2(v)) = H_1(u) \cdot H_2(v)/2m + 0.5$, since $H_1(u)$ and $H_2(v)$ are binary vectors. Without loss of generality, assuming $f(v,u) \in [0,1]$, this can be thought of as a problem of inner product fitting with binary codes computed through the two hash functions. This formulation essentially  establishes an asymmetric hashing scheme because the hash function $H_1$ is applied on the user (query) while a different hash function $H_2$ is applied on the items (in the database). These hash functions are jointly learned so that their discrete inner product fits the similarity structure defined by the binary function $f$. Note that this is different from asymmetric MIPS algorithms, e.g., ~\citet{shrivastava2014asymmetric}, which are linear and are designed for some specific non-metric measures. In contrast, via inner product fitting, FLORA can learn a pair of non-linear hash functions whose inner product, and equivalently Hamming distance, approximately reveals the true similarity given by any definition of $f$.}

{The discrete-optimization objective in Eq.~\eqref{eqn:consistency_objective} is computationally intractable to solve. Therefore, we replace the discrete optimization with a continuous relaxation where the last layers of the hash functions approximate the discrete constraint using the $tanh$ activation. Specifically, given the continuous function $h_1: u \rightarrow [-1,1]^m$ and $h_2: v \rightarrow [-1,1]^m$ whose last layers are with the $tanh$ activation, the consistency loss is as follows:}
\begin{equation}
    \mathcal{L}_c  = E_{u, v} ||f(v, u) - cosine(h_1(u), h_2(v))||_2
    \label{eqn:consistency_objective_relaxed}
\end{equation}

Besides the consistency objective, the model also needs to learn balanced hash codes for optimal searching~\citep{doan2020efficient}. First, we need to prevent collapsed bits. This happens when the training algorithm fails to learn a specific bit, and thus it assigns most examples into the same bit value. Intuitively, this means that the data points should be uniformly distributed into the hash codes in the discrete space. To achieve balanced codes, we propose to solve two additional objectives. The first term $L_u$ encourages half of the points to be assigned to either $-1$ or $1$, and can be defined as follows:
\begin{equation}
    \mathcal{L}_u  = \sum_{k=1}^m \big| \sum_{i=1}^{n_1} h_{1,k}(u_i) \big|_1 + \big| \sum_{j=1}^{n_2} h_{2,k}(v_j) \big|_1
    \label{eqn:uniform_objective}
\end{equation}
where $h_{1,k}$ and $h_{2,k}$ represent the $k$th output components of $h_1$ and $h_2$, respectively. The final objective discourages different bits to learn correlated information of the data. Essentially, the projection of the data points onto the discrete space should be an orthogonal projection. This objective, denoted as $\mathcal{L}_i$, is defined as follows:
\begin{equation}
    \mathcal{L}_i  = ||W_{h_1}^T W_{h_1}^{} - I||_2 + ||W_{h_2}^T W_{h_2}^{} - I||_2
    \label{eqn:independent_objective}
\end{equation}
where $W_{h_1}$ and $W_{h_2}$ are the parameter matrices at the last layers of the hash functions $h_1$ and $h_2$, respectively. Since $h_1$ and $h_2$ share their last layers, we have $W_{h_1} =  W_{h_2}$. The overall objective function of FLORA is a combination of the consistency loss, uniform frequency loss, and independent-bit loss, as follows:
\begin{equation}
    \mathcal{L}  = \mathcal{L}_c + \lambda_u \mathcal{L}_u + \lambda_i \mathcal{L}_i \label{eqn:overall_objective}
\end{equation}
where $\lambda_u$ and $\lambda_i$ are parameters to balance the losses. Optimizing $\mathcal{L}$ learns two hash functions $h_1$ and $h_2$ that generate balanced hash codes and together preserve the similarity structure defined by the binary function $f$ in an asymmetric hashing scheme.

\subsection{Hash Function Training}

Training the model in FLORA is not a trivial task, mainly because  $\mathcal{D}_{orig}$ is generally not available. There are several challenges when the training process can only access the binary function (possibly without the knowledge of its internal structure) and the input features of the two domains. The number of possible training input pairs, i.e., $n_1 \times n_2$, can be extremely large, which requires a significant amount of training time and computation. Furthermore, in a typical recommendation or retrieval setting, the number of ``relevant'' items for each user is often very small. Consequently, a large number of user-item data pairs do not contain useful signals to train the model and can even make the training process more difficult to converge into a good local optimum.

\begin{figure}[h]

    \centering
    \includegraphics[width=6in]{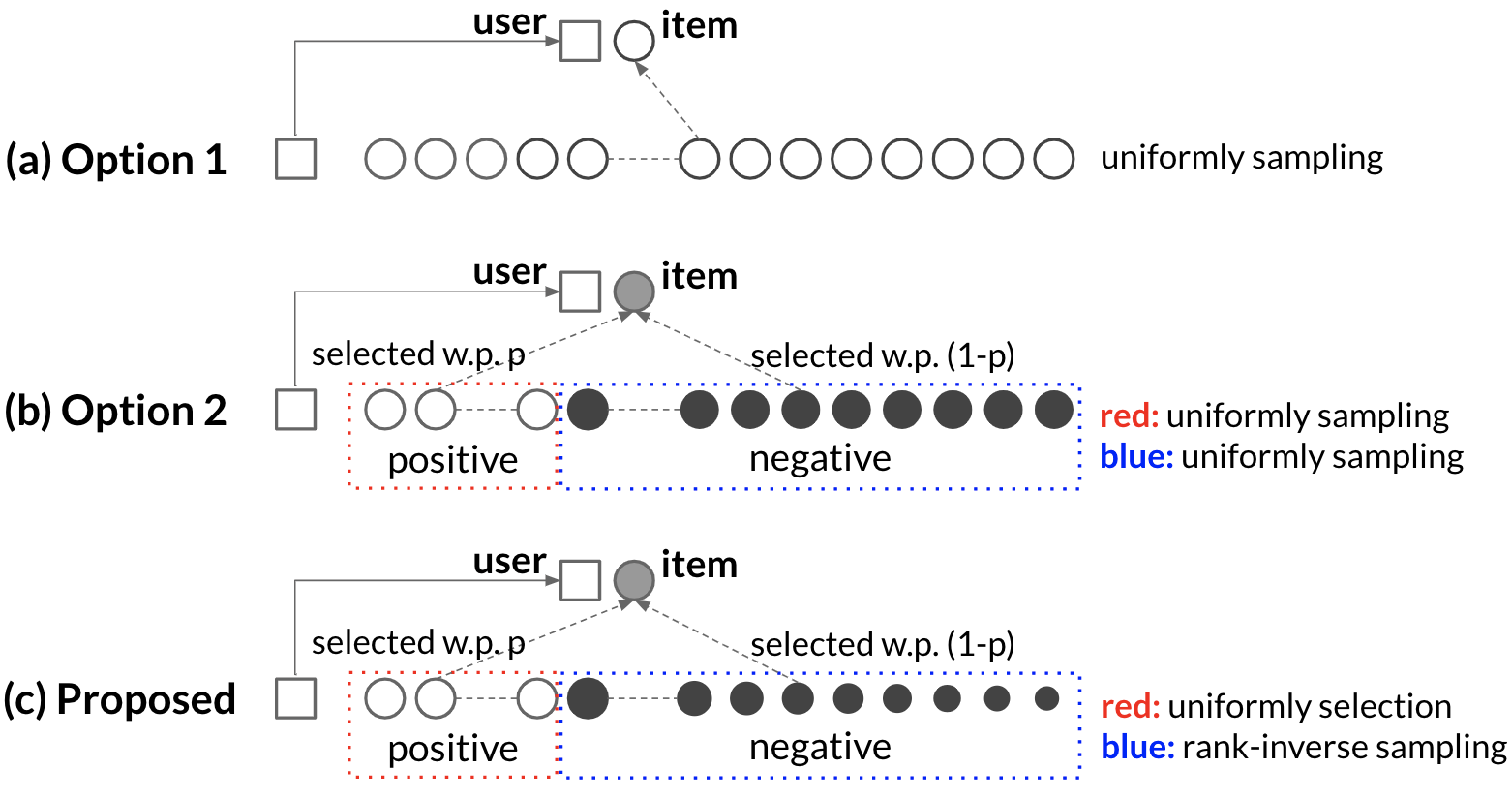}
    \caption{FLORA's sampling strategy for model training, compared to other possible sampling strategies.}
    \label{fig:fltr_sampling}
\end{figure}

It is crucial that the FLORA framework needs an efficient training approach, i.e., an efficient sampling of $\mathcal{D}_{app}$. One natural option is to randomly sample the pairs for training, as depicted in Option 1 in Figure~\ref{fig:fltr_sampling}(a). However, as later shown in Section~\ref{sec:exp}, this procedure does not work well. To achieve good retrieval performance, the hash functions need to agree on the relevant, positive user-item pairs while disagreeing on the negative user-item pairs. Randomly sampling the pairs does not increase the chance of the relevant data pairs being selected, thus it may take a large number of training steps to adequately iterate over these relevant pairs. Inspired by this observation, we propose two efficient sampling strategies.

\newpage

Let $N_p$ be the number of top $N_p$ items (w.r.t the original binary function $f$) of a user. In training, for a user, we select an item to form the user-item input as follows: with a probability $p$, we select an item from the top $N_p$ items (called the positive set), and with probability $1-p$, we select an item from the set of remaining items (called the negative set). Generally, $p$ is set to $0.5$ to allow uniform selections of the positive and negative items, thus, hereafter, we assume $p=0.5$. Note that, $N_p$ is small for real-world applications, where a user is typically interested in a small number of items. Given this sampling setup, selecting an item from the top $N_p$ items is trivial because $N_p$ is small. Sampling from the negative set can be done uniformly (illustrated in Option 2 in Figure~\ref{fig:fltr_sampling}(b)).

While this strategy provides some noticeable improvement over Option 1, we empirically show that the second proposed strategy (Option 3) results in even faster training convergence and a significantly better fast ranking performance. In Option 3 (illustrated in Figure~\ref{fig:fltr_sampling}(c)), a negative item is sampled as follows: an item in the negative set is selected with a probability inversely proportional to their rank in the set. There are two reasons that this strategy  results in faster training and better performance than the other procedures. First, items on the top of the negative set's ranked list have more chance of being selected; these items are more likely to provide a relevant signal to train the model than the items near the end of the ranked list. Second, items near the end of the ranked list still have a chance of being selected; these negative items can provide useful signals for the model to push the hash codes of relevant items of a user away from those of irrelevant items during training.

For each proposed sampling option, the pairs are repeatedly sampled to form the minibatches. Then, FLORA's hashing model is trained on the sampled minibatches with stochastic optimization.

\subsection{Fast Ranking with FLORA}

After the two hash functions are learned, ranking in FLORA is conducted in the learned discrete space using an asymmetric hashing scheme. The hash codes of the query $q$ and the item $x$ are $H_1(q)=sign(h_1(q))$ and $H_2(x)=sign(h_2(x))$, respectively. The hash code of the query is typically computed online via $H_1$, while the items' hash codes can be pre-computed using $H_2$ and serve as an indexing structure (i.e., the hash table) for fast retrieval. {There are two strategies to perform fast retrieval of the items. Given a query, the first strategy is to enumerate the hash codes within a Hamming ball of the query and use them for probing candidate items with the same hash codes in the hash table. This works well when there exist candidate items within a short Hamming distance to the query; otherwise, a large number of hash codes must be enumerated in order to probe candidate items. }

The second strategy leverages the fact that the hash codes are low-dimensional, discrete objects and can be operated efficiently in both computation and memory. The size $m$ of the hash codes is often between 32 to 128 bits (equivalent to the size of 1 to 3 integers, respectively). Therefore, the entire set of pre-computed, discrete hash codes of the items can be efficiently kept in memory (e.g., a million 32-bit hash codes result in 4MB of memory), and the distance calculation can be extremely fast (i.e., using one XOR and one bit-count operation, each of which typically takes 1 CPU cycle, to compute the Hamming distance). The approximate ranking is performed via the Hamming distances calculated between the query's hash code and the set of pre-computed items' hash codes. As we shall show later, this ranking is fast and achieves a superior approximation of the exact ranking using the original binary function on all datasets. Note that with other approaches, especially the  SL2G method~\citep{tan2020fast}, it becomes prohibitively slow when the binary neural function $f$ is complex and the required number of retrieved items grows because each traversal in these ANN index structures requires a more expensive computation on the binary function.

\newpage

\section{Experiments}
\label{sec:exp}

We evaluate the performance of FLORA in the fast ranking task under different combinations of generic searching measures $f$'s and datasets in the recommendation and advertising domains. It is important to note that the comparison of the neural network-based recommendation models (i.e., $f$) is beyond the scope of this paper. The measures $f$’s are pre-trained on the respective datasets, and our evaluation only focuses on the fast item ranking~procedure.

\subsection{Neural Network-based Ranking Measures}

\begin{figure*}[h]
    \centering
    \includegraphics[width=6.7in]{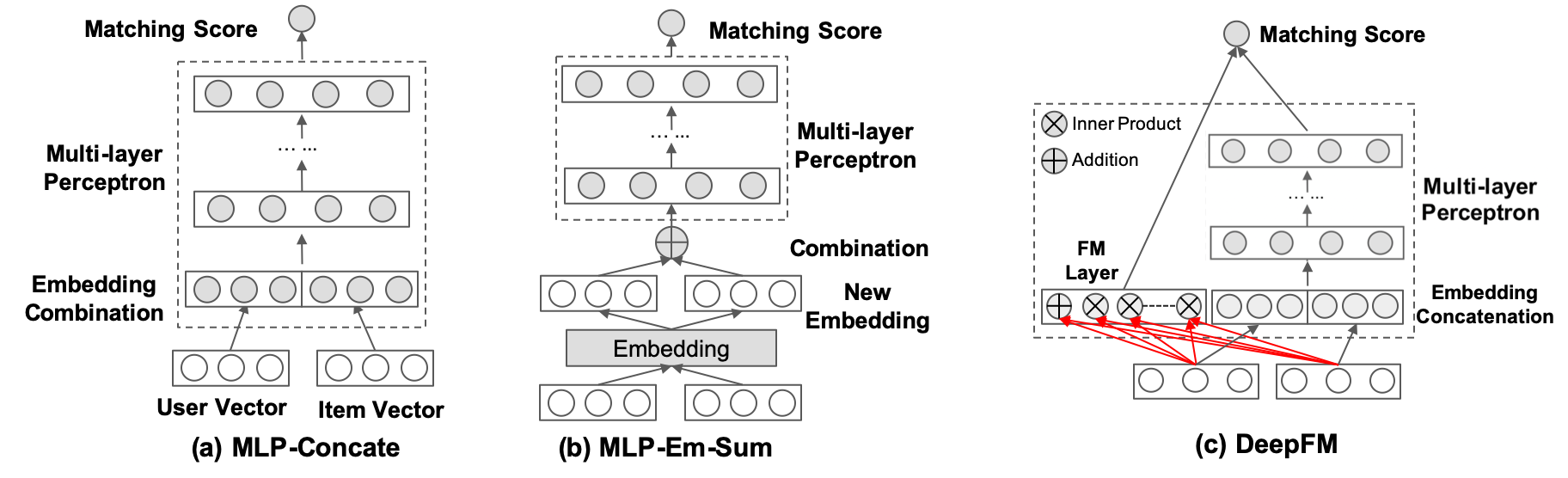}
    \caption{Two variants of the MLP models for recommendation and a Deep\&Wide variant, DeepFM, also for recommendation. 
    }
    \label{fig:neural_ranking_models}
\end{figure*}

We leverage three popular neural network  models in recommendation and advertising. The first two MLP models, \textbf{MLP-Concate}~and \textbf{MLP-Em-Sum}, were used in~\citep{tan2020fast} to evaluate~the fast neural ranking methods. MLP-Concate concatenates user latent vectors and item latent vectors before they go through the MLP network (Figure ~\ref{fig:neural_ranking_models} (a)).  MLP-Em-Sum transforms the heterogeneous vectors from the two domains into a common embedding space before the merge operation (Figure~\ref{fig:neural_ranking_models} (b)). Consequently, user vectors and item vectors will lie on the same manifold on which element-wise sum will be operated. The third model, \textbf{DeepFM}~\citep{guo2017deepfm}, adopts the popular Wide\&Deep architecture with a separate factorization-machine layer and an MLP. DeepFM has been shown to be able to learn low and high-order feature interactions simultaneously from the raw input and demonstrates good performance in both  recommendation and click-through-rate prediction~tasks.

\subsection{Datasets}

We choose three widely used datasets for recommendation: \textbf{Yelp}\footnote{\url{https://www.Yelp.com/dataset/challenge}}, AmazonMovie\footnote{\url{http://jmcauley.ucsd.edu/data/amazon}} (\textbf{Amovie}) and Movielens-20M\footnote{\url{https://grouplens.org/datasets/movielens/20m}} (\textbf{Movielens}). The Yelp dataset is pre-processed and contains 25,677 users, 25,815 items, and 731,670 ratings. For Amovie, we remove the users with less than 30 interactions. This results in 7,748 users, 104,708 items, and 746,397 ratings. Finally, for Movielens, we randomly select 25,000 users from the original dataset of 20 million ratings, created by 138,000 users on 27,000 movies. The resulting Movielens dataset contains 18,799 movies, and 3,670,197 ratings. Note that the number of ratings in Movielens is significantly larger than those in Amovie and Yelp. There are also more ties on top ranking (w.r.t $f$) in Movielens than in the other datasets. Note that, the rating data constitute the original pairwise dataset $\mathcal{D}_{orig}$ that is used to train $f$.


We train MLP-Concate and MLP-Em-Sum on Yelp and Amovies, and DeepFM on Movielens to learn the corresponding neural functions. After training, we freeze the weights of the models; the frozen models become the binary function $f$’s. For each pair of user and item input vectors, the binary function $f$ outputs a similarity score between 0 and 1. Note that, for all models, only the networks above the vector combination (i.e., the part within the dashed boxes in Figure~\ref{fig:neural_ranking_models}) are regarded as the matching function. The dimensions of the input vectors for MLP-Em-Sum, MLP-Concate and DeepFM are set as 64, 32, and 100, respectively.


Note that after training $f$, to ensure a general OBFS setting, we assume that the dataset $\mathcal{D}_{orig}$ becomes unavailable. Only the two separate sets of queries and database items (i.e., users and items, respectively) in each of the three datasets are available to train the model or build the ranking index in the OBFS solutions. In other words, sampling the pairs of users and items from these two separate sets induces $\mathcal{D}_{app}$. For each dataset, we first randomly sample 1,000 users for testing. The remaining users are used for learning the hash functions.


\subsection{Comparison Methods}

We compare FLORA against two state-of-the-art methods for traditional ANN search, ANNOY and HNSW~\citep{malkov2020efficient,zhao2020song}, the  hashing-based recommendation approach, CIGAR~\citep{kang2019candidate}, that uses pairwise/implicit feedback data for candidate ranking, and the current state-of-the-art for fast neural ranking, SL2G. ANNOY mainly focuses on metric measures. Similar to~\citep{tan2020fast}, we first retrieve the candidates by ANNOY via $\ell_2$-distance (here, user vectors and item vectors are required to have the same dimension), then re-rank the candidates via the searching measure $f$. We modify HNSW for neural-network binary functions as follows: we calculate $f(x, x)$ for a pair of item vectors forcibly. Then, we use the value of $f(x, x)$ as the relevance measure to construct the index graph by HNSW algorithm. This index-graph construction may be dramatically different from the Delaunay graph with respect to $f(x, q)$; thus, there is no performance guarantee. Note that we use the same dimension for the users and items so that the modified HNSW can be implemented; otherwise, $f(x, x)$ cannot be computed. CIGAR trains the candidate-ranking hash function using the original pairwise rating dataset (i.e., $\mathcal{D}_{orig}$). However, since we do not have such a dataset in OBFS, we first randomly sample a pair of users and items and calculate the corresponding rating using $f$. This process essentially produces $\mathcal{D}_{app}$, which is used to train the CIGAR's hash functions. SL2G and FLORA have no such limitations.

\vspace{0.1in}

Note that it is not possible to evaluate most hashing approaches~\citep{gionis1999similarity,gong2013iterative,shen2015supervised,yang2018supervised,cao2018hashgan,huang2017unsupervised,lin2016learning,huang2016unsupervised,salakhutdinov2009semantic} in this work since these methods are designed for specific measures, e.g., $\ell_2$, instead of arbitrary measures such as neural networks. Evaluating cross-modal hashing approaches~\citep{jiang2017deep,li2018self,zhang2018attention,gu2019adversary,kang2019candidate} is also very difficult since the required $\mathcal{D}_{orig}$ is not available.  Nevertheless, we provide comparisons with CIGAR~\citep{kang2019candidate}, a representative cross-modal retrieval method that can perform candidate ranking by learning a hash function.

\newpage

\noindent \textbf{Implementation Details}: The asymmetric hashing network in FLORA is implemented with Multilayer Perceptrons (MLPs). Each user or item  tower is an MLP with two layers (256-256). The shared MLP consists of two layers of 128 and $m$ units, where $m$ is the dimension of the hash codes. Typically, higher numbers of bits result in better retrieval performance but also increase the computation in both training and retrieval; thus we set $m=128$, which has been previously shown to achieve the best retrieval performance with acceptable computational efficiency~\citep{cao2018hashgan,huang2017unsupervised,lin2016learning,huang2016unsupervised,salakhutdinov2009semantic,zhang2018attention,gu2019adversary,kang2019candidate}. We use small grid search in $\{0.1, 1, 10\}$ for both $\lambda_u$ and $\lambda_i$ and set the optimal value for each dataset using a corresponding validation user set (10\%  of the training users).

\subsection{Evaluation Metrics}


We first generate the ground-truth labels by calculating the most relevant items for each user via the learned binary function $f$. Similar to~\citet{tan2020fast}, we perform experiments on Top-10 and Top-100 labels. {In Top-10 (or Top-100), an item is relevant to the query if it is one of the top 10 (or top 100) items in the ground-truth ranking computed using $f$}. Since we focus on the fast ranking task, the main evaluation metric is recall; we do not evaluate the methods on the precision. The recall values are calculated up to 200 retrieved items, a typical range for the ranking tasks in real-world applications.

Since fast ranking with our hashing approach is very different from the existing ANN searching algorithms, suitable evaluation approaches should be used. If we compare FLORA against the baselines in metrics such as Recall versus Time (i.e., how many times the algorithm can speed up over naive brute force scanning at each recall level), FLORA will have an unfair advantage over the baselines since each distance computation in FLORA (in the discrete space) is significantly faster than each distance computation in the baselines (via the binary function $f$). Therefore, we focus our evaluation of the methods on their ranking qualities via the recall performance. This way, when FLORA achieves good recall performance, we can confidently conclude that FLORA is significantly fast and produces a better approximate ranking for the OBFS problem with $f$.

\subsection{Ranking Performance Results}

\begin{figure*}[t!]


\centering
\mbox{
		\includegraphics[width=2.8in]{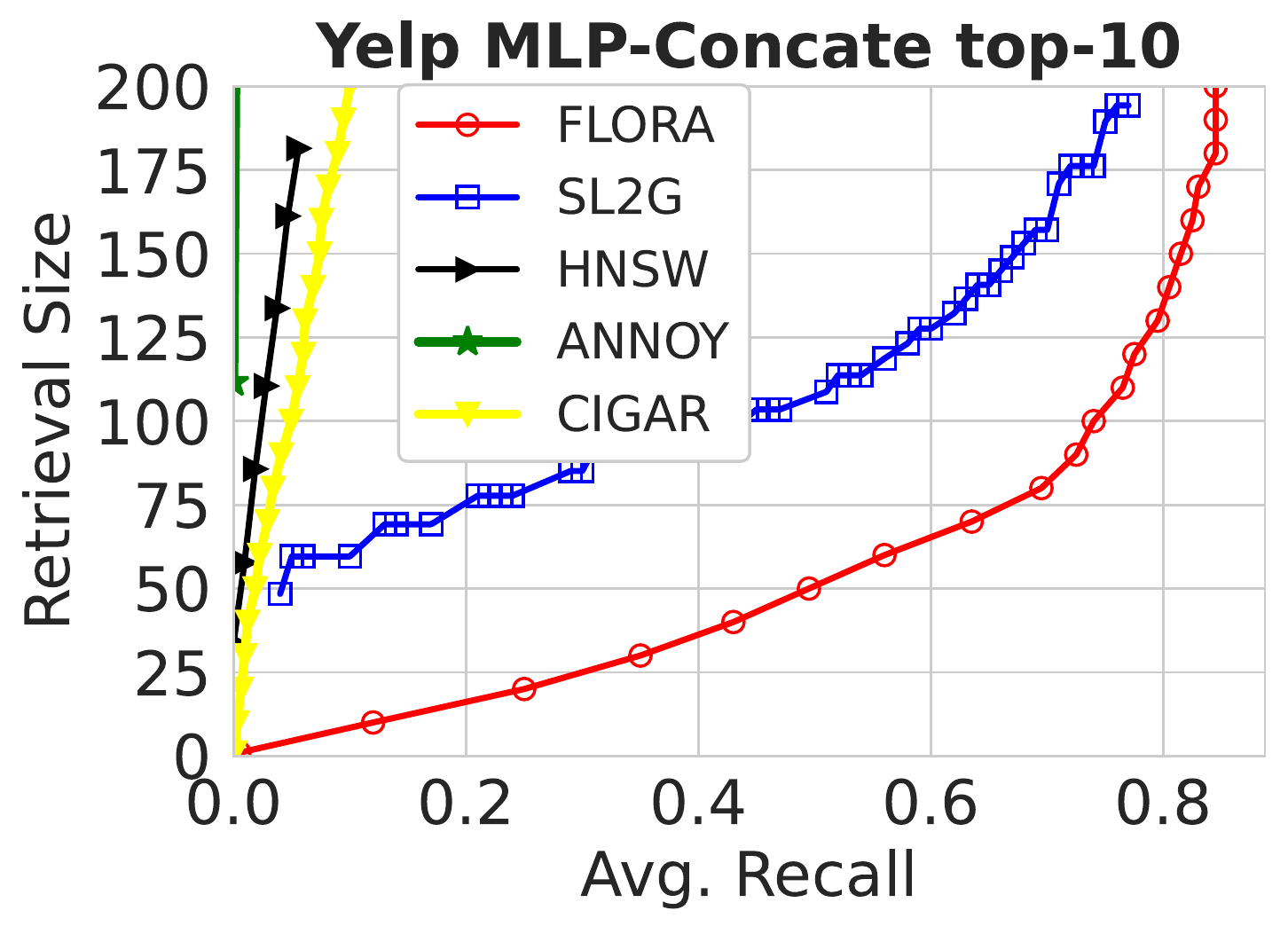}\hspace{0.2in}
		\includegraphics[width=2.8in]{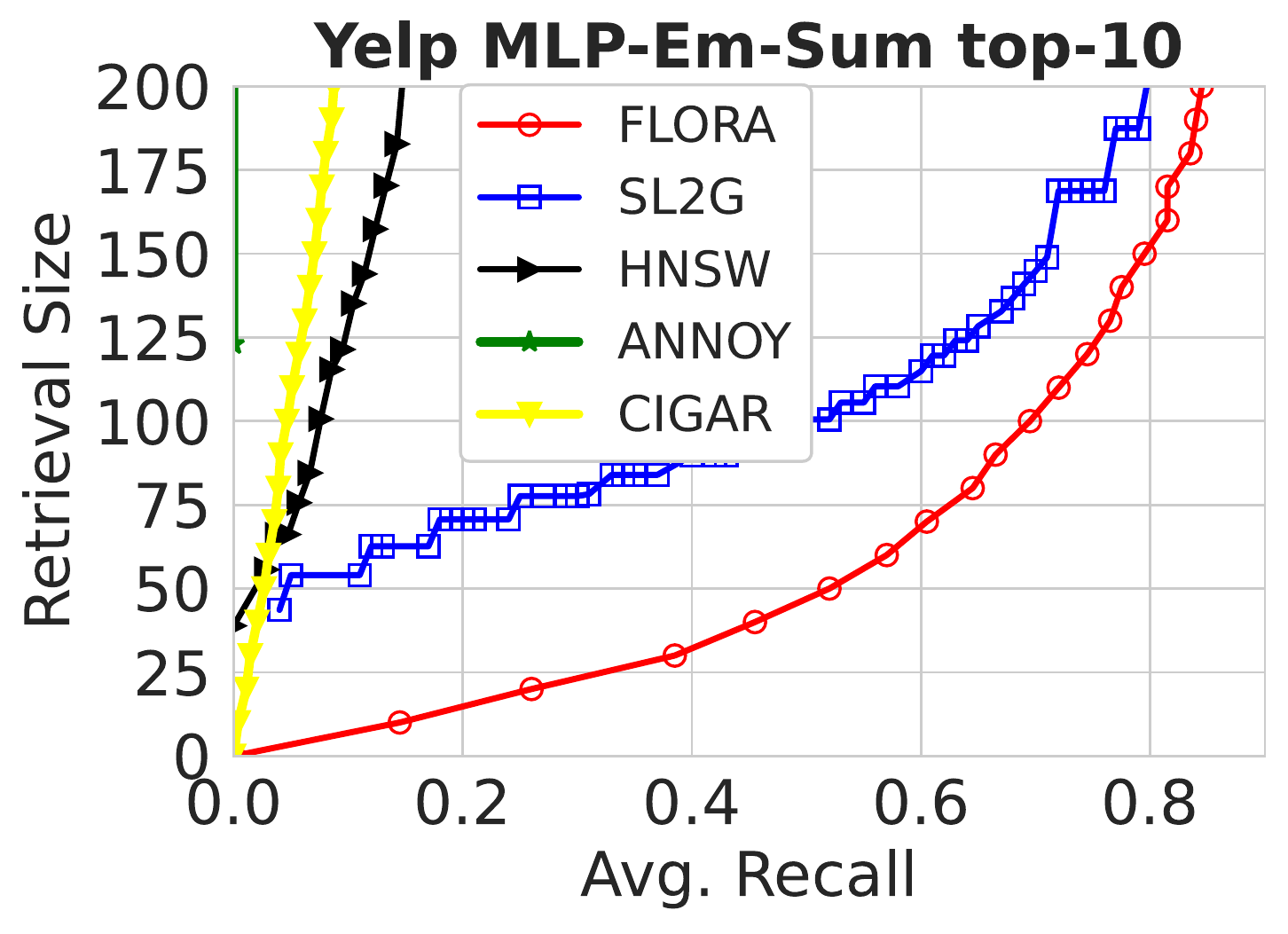}
}

\mbox{
		\includegraphics[width=2.8in]{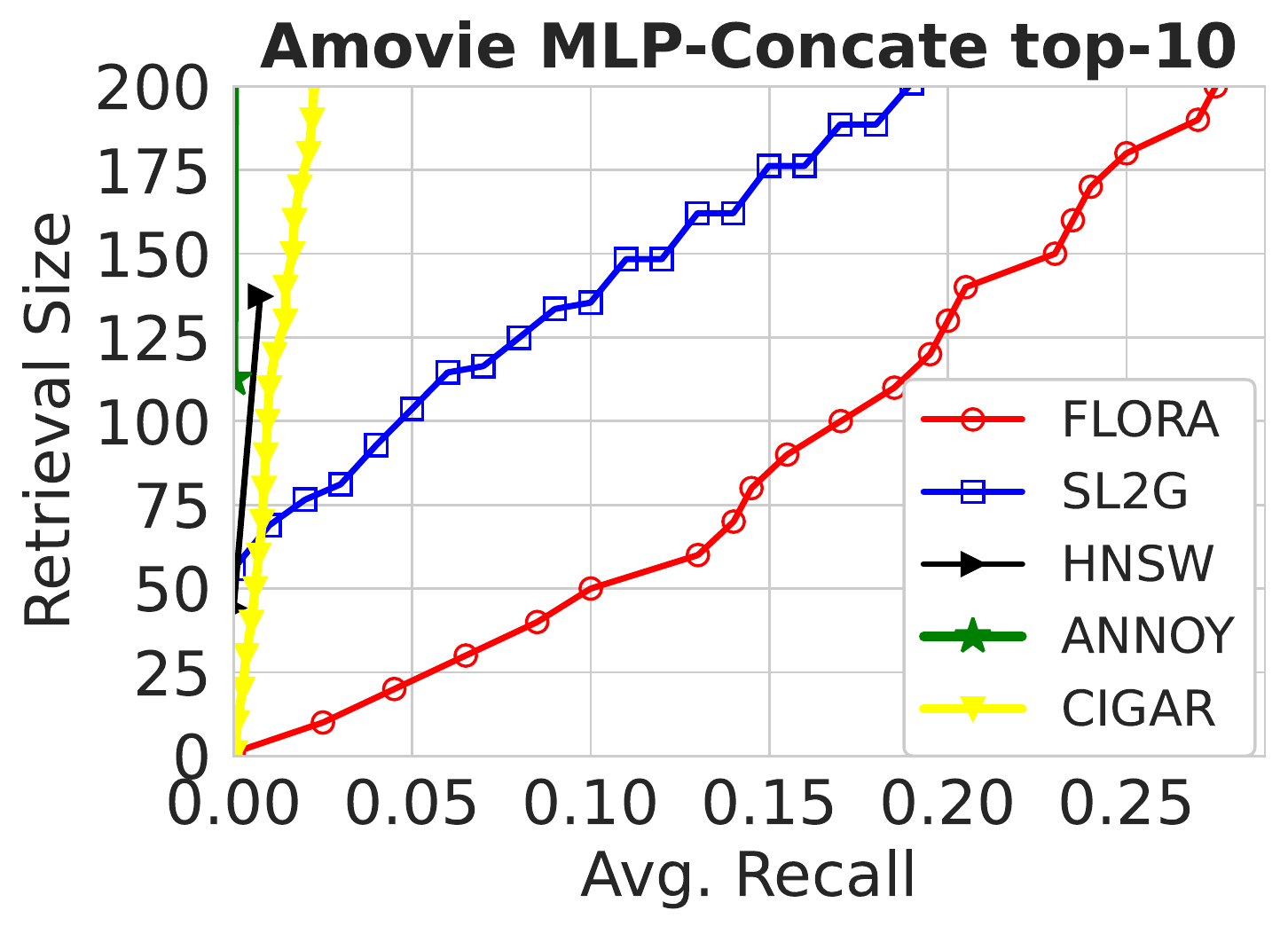}\hspace{0.2in}
		\includegraphics[width=2.8in]{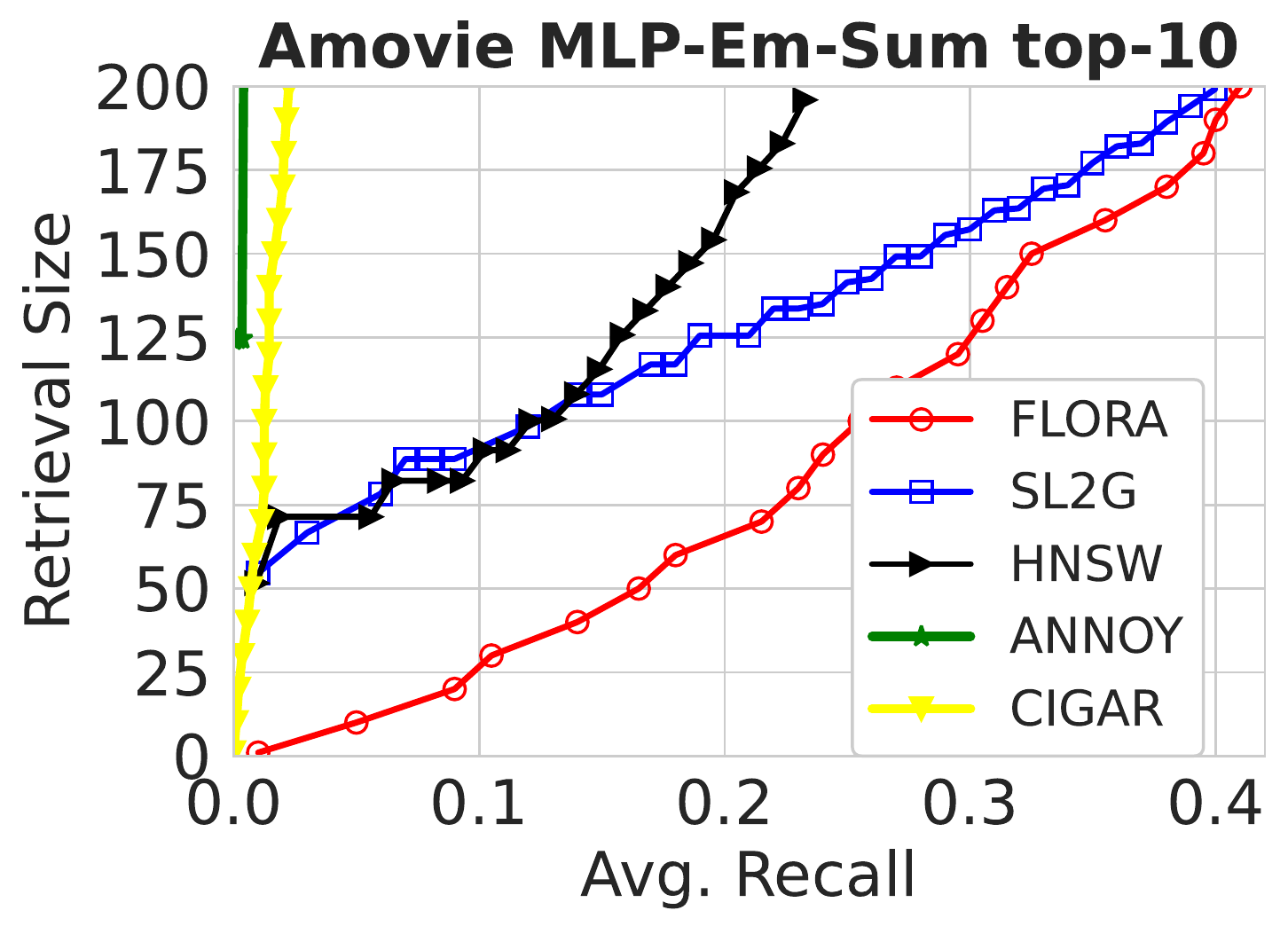}
}
    \vspace{-0.1in}
	\caption{Top-10 recall results at different retrieval thresholds for MLP-Concate and MLP-Em-Sum.}\label{fig:results_retrieval_recall_top10}
\end{figure*}

\begin{figure}[t!]	
\centering
\mbox{
		\includegraphics[width=2.8in]{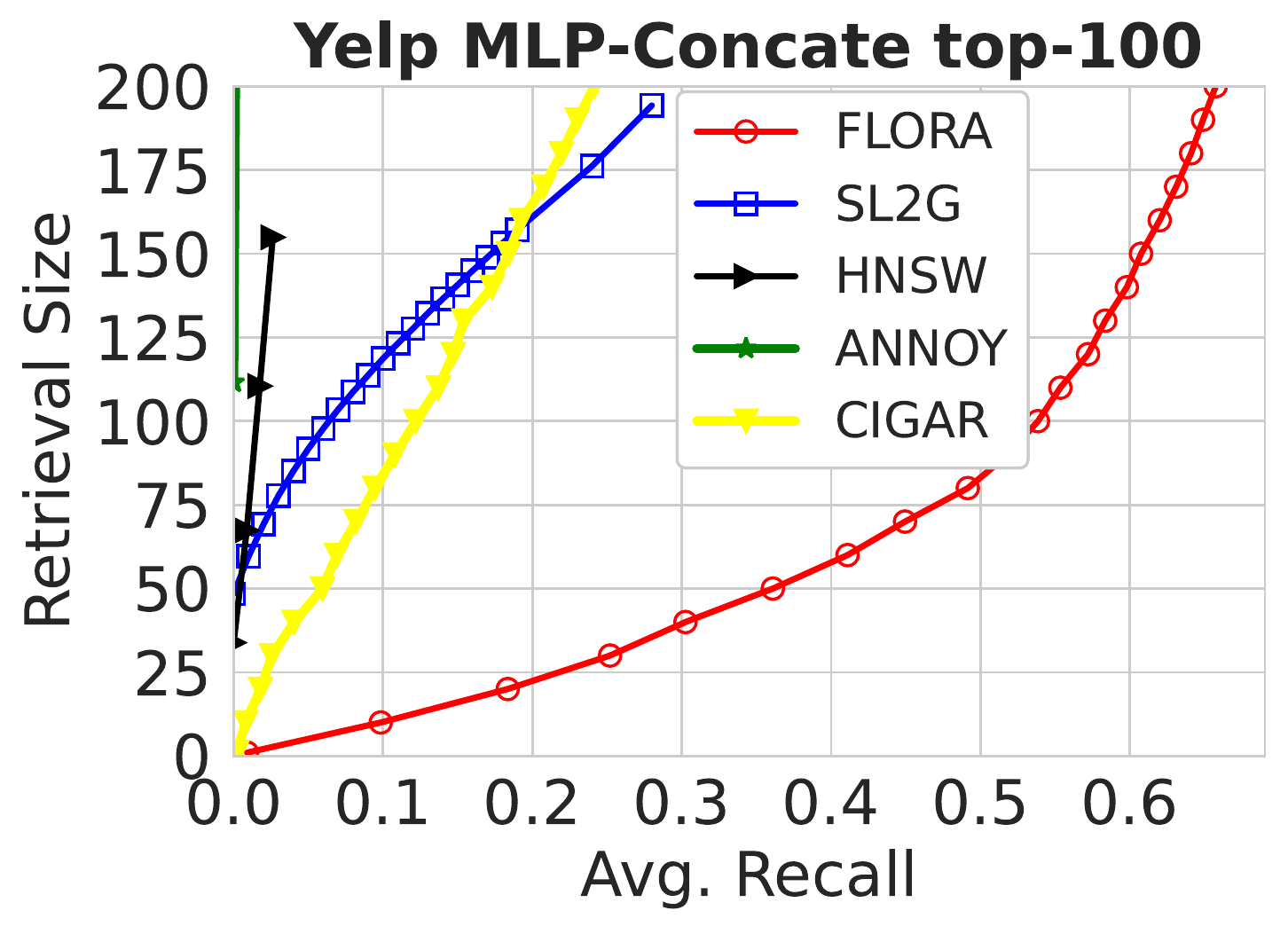}\hspace{0.2in}
		\includegraphics[width=2.8in]{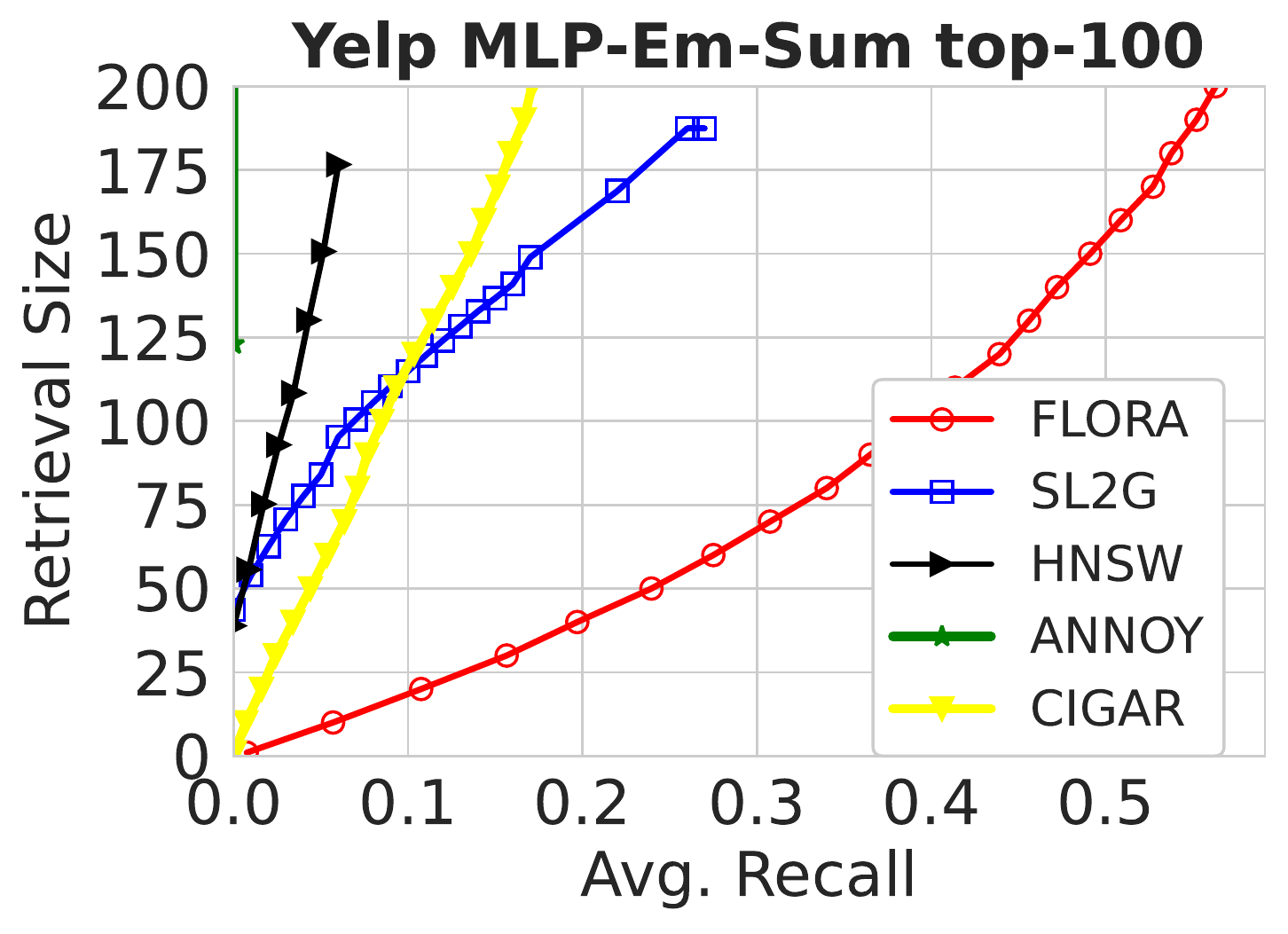}
}

\mbox{
		\includegraphics[width=2.8in]{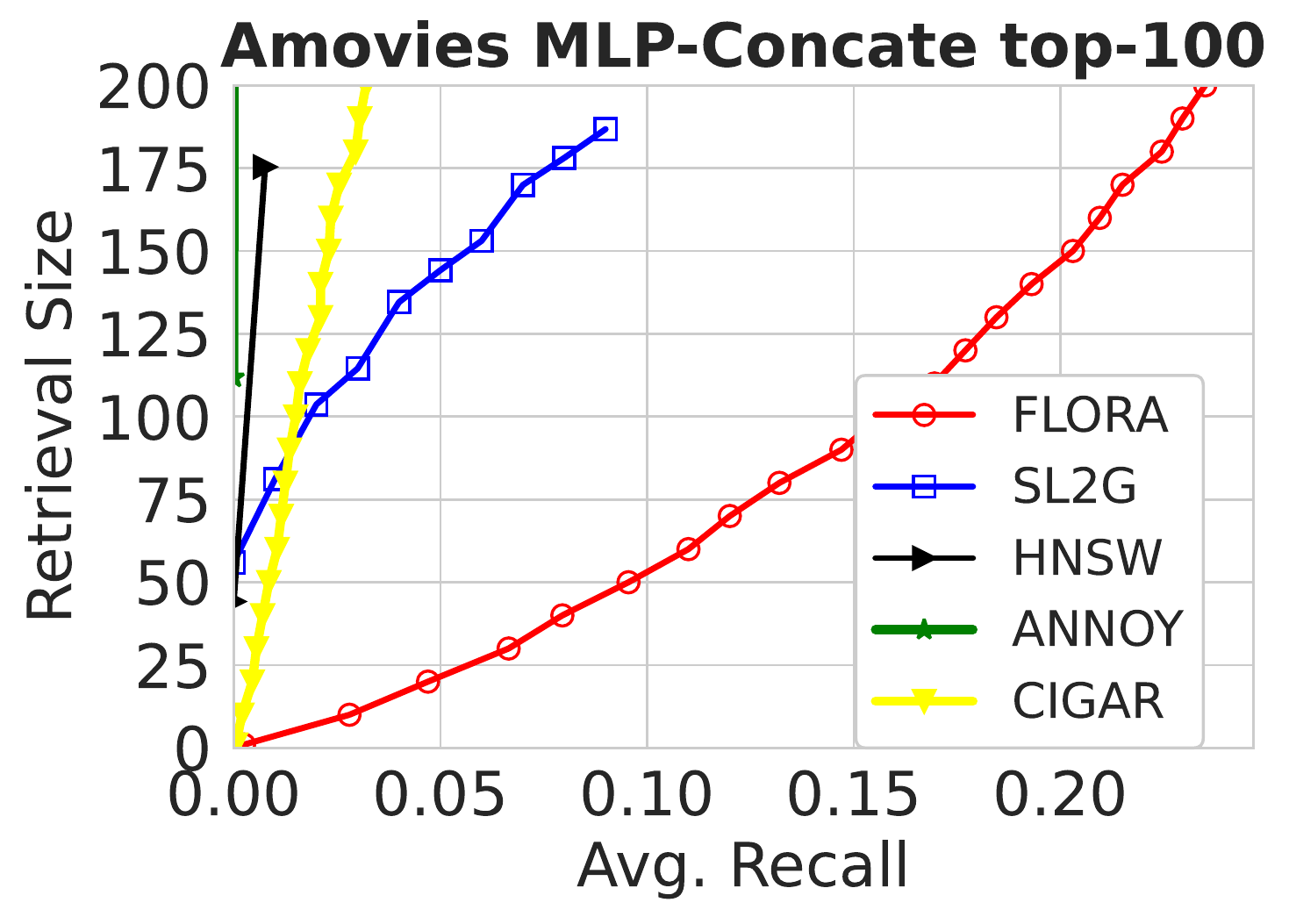}\hspace{0.2in}
		\includegraphics[width=2.8in]{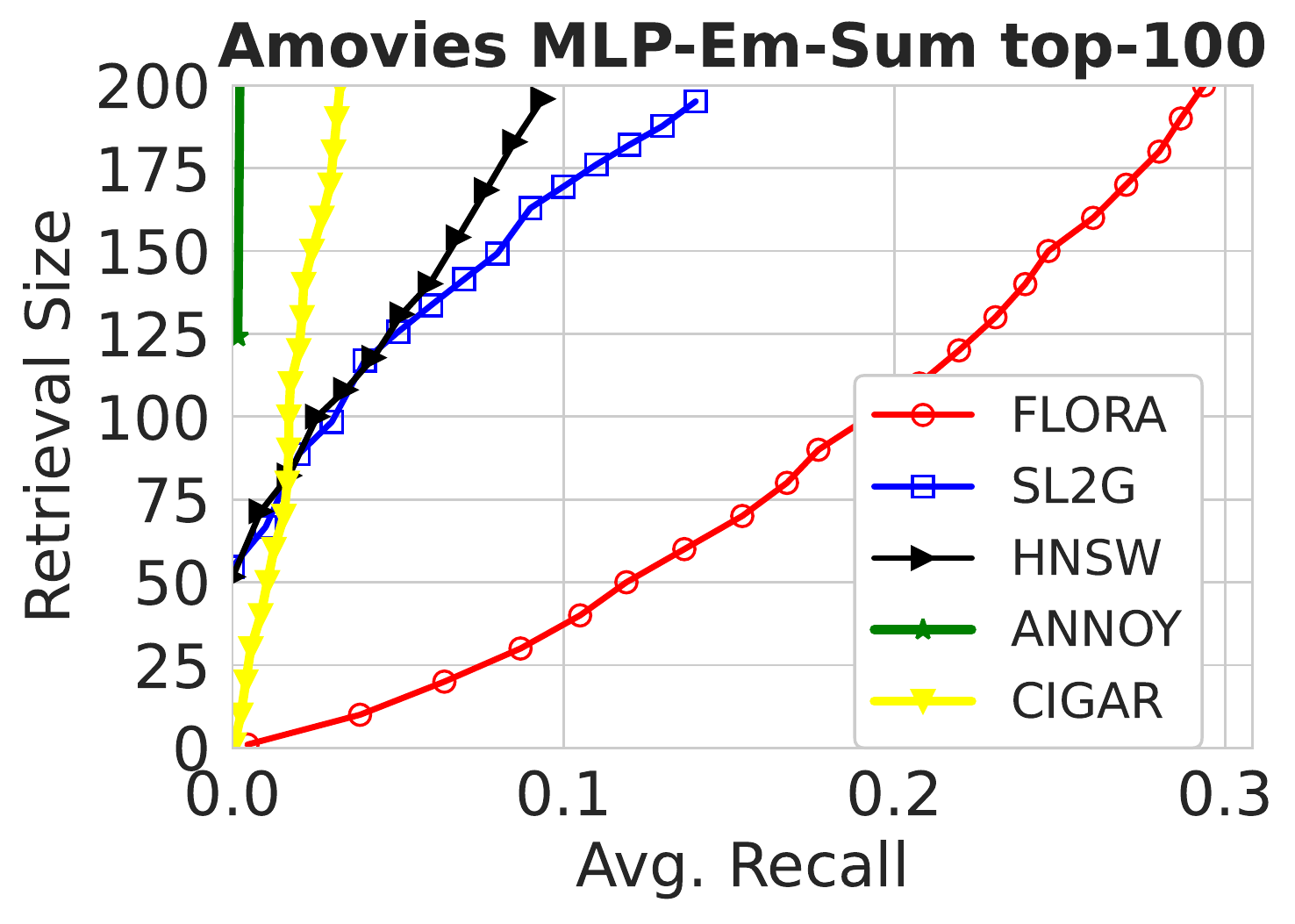}
}
    \vspace{-0.1in}
	\caption{Top-100 recall results at different retrieval thresholds for MLP-Concate and MLP-Em-Sum. }\label{fig:results_retrieval_recall_top100}\vspace{-0.1in}
\end{figure}

\begin{figure}[h]
\centering
\mbox{
    \includegraphics[width=2.8in]{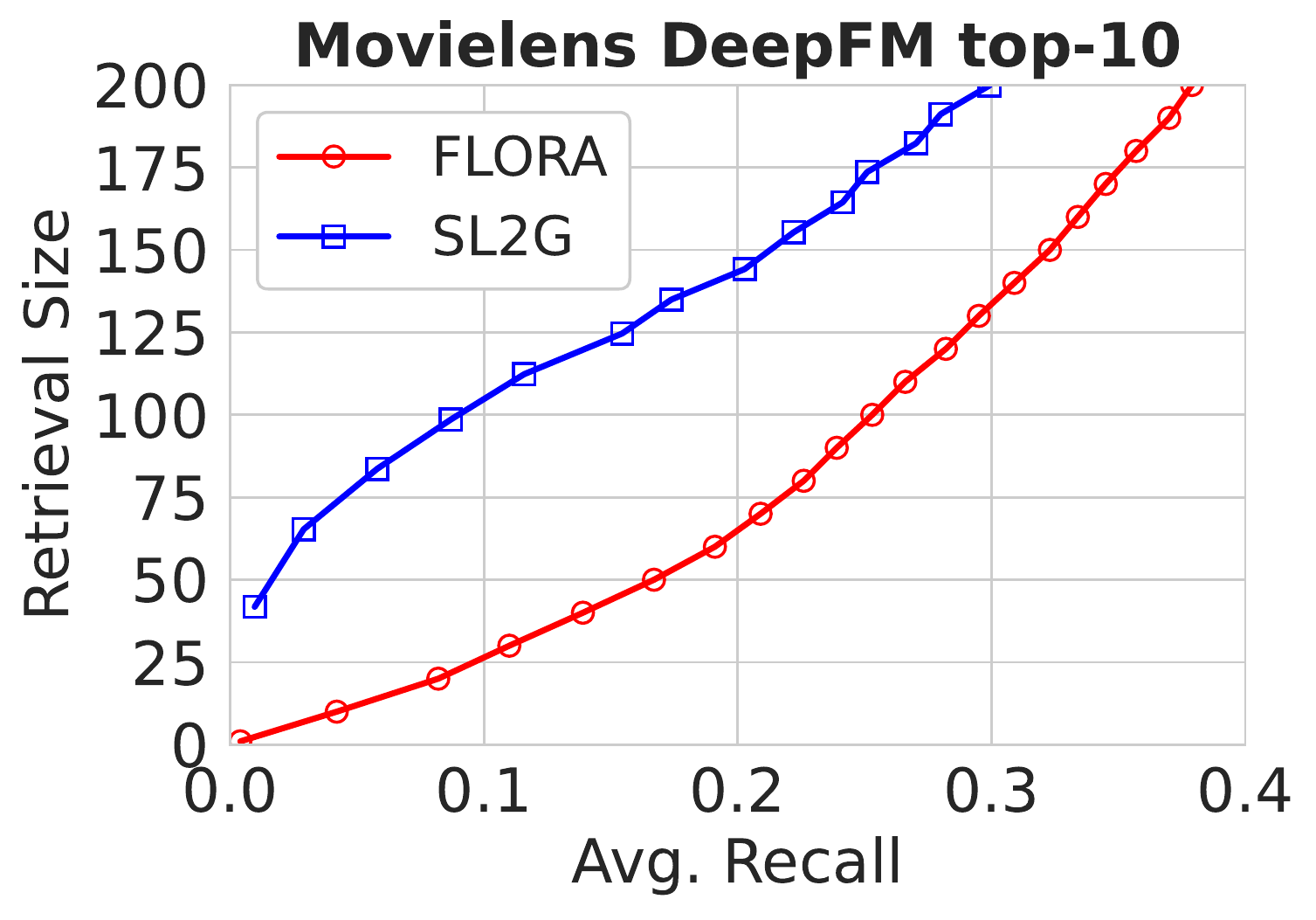}\hspace{0.2in}
	\includegraphics[width=2.8in]{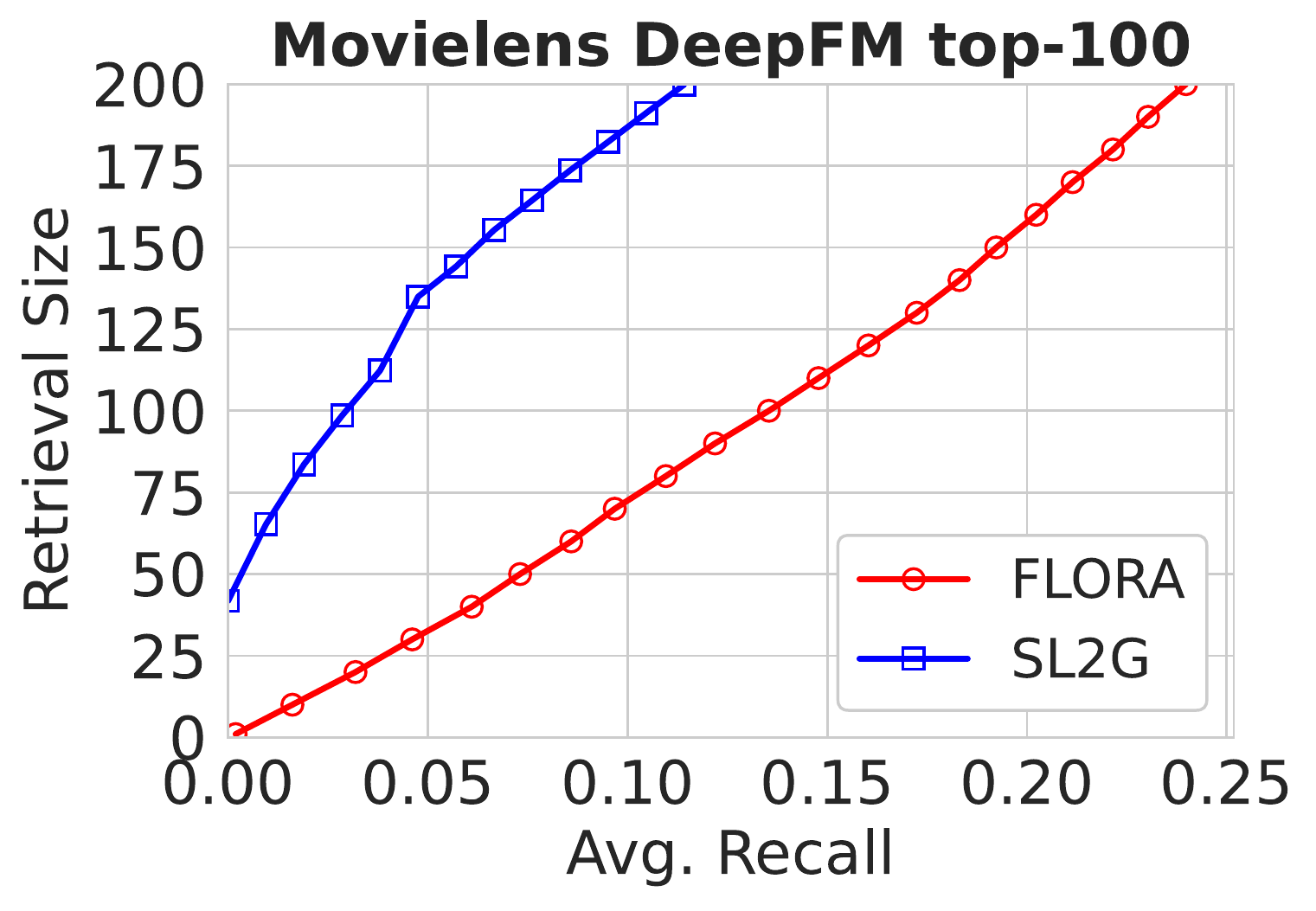}
}
\vspace{-0.1in}
	\caption{Top-10 and Top-100 recall results at different retrieval thesholds for DeepFM on Movielens. }\label{fig:results_retrieval_recall_movielens}
\end{figure}

We show the recall performance at various retrieval thresholds of the methods for the Top-10 and Top-100 retrievals on Yelp and Amovies datasets in Figure~\ref{fig:results_retrieval_recall_top10} and Figure~\ref{fig:results_retrieval_recall_top100}, respectively. Additionally, we present the Top-10 and Top-100 results for SL2G and FLORA on the Movielens dataset in Figure~\ref{fig:results_retrieval_recall_movielens}. As can be seen, both ANNOY, a metric-measure ANN method, and the modified HNSW are not able to retrieve the most relevant items. CIGAR's retrieval performance is also significantly worse than FLORA's performance.


{Previously, \citet{tan2020fast} show that SL2G achieves both significantly better ranking performance, which we also observe in this paper, and computational speed compared to these ANN methods.} However, FLORA consistently outperforms SL2G in the experiments on all datasets, especially at the top of the ranking list.

\newpage

Next, we present some observations, along with important conclusions about the difference between FLORA and other methods:
\begin{itemize}
    \item Existing hashing-based methods are not suitable to solve the OBFS problem. The main challenge in adapting these methods is their training efficiency when we do not have access to $\mathcal{D}_{orig}$ and have to rely on the huge $\mathcal{D}_{app}$, which consists of all possible user-item pairs. For example, the sizes of the enumerated  $\mathcal{D}_{app}$ for Yelp, Amovie, and Movielens are approximately 662 millions, 811 millions and 469 millions, respectively.  Later, we empirically demonstrate how FLORA overcomes this challenge with the proposed sampling approach in Section~\ref{sec:result_sampling_efficiency}.
    \item FLORA exhibits significantly better early-rise recalls compared to SL2G, even though it performs the fast ranking in the compact, discrete space. As discussed, ranking in the discrete space is extremely fast, where each distance calculation is hundreds to thousands of times faster than the computation on the original binary function $f$. The item ranking can also be performed in memory. This is not possible with the compared baselines.
    \item  While FLORA's recall curves approach those of SL2G at the end of the ranking lists in a few cases, the superior results of FLORA are still significant as the preferred retrieval sizes in real applications are typically no more than a few hundred items (also for such reason, we select 200 in the experiments). Furthermore, since FLORA is based on hashing, we can systematically improve its recalls  by employing multiple hash tables, as will be discussed~later.

    \item When the binary function is DeepFM, we observe that the ranking computation of SL2G, which involves several iterations over the binary function to compute the distances, becomes too slow to be practical. In contrast, FLORA's ranking computation, performed in the discrete space, is nearly constant in all experiments.
\end{itemize}

\newpage

\subsection{Exact Retrieval with Neural Function Re-ranking}

FLORA ranks the items in the discrete space. This procedure has a trade-off: more items than required are sometimes returned if there are ties at the specified threshold in the ranking list. Suppose we want to retrieve only 1 item for a user; FLORA may instead return multiple candidate items if these items have the same shortest Hamming distance to the user. Therefore, a re-ranking step is needed to retrieve exactly 1 item. The re-ranking step can be performed using $f$. Note that invoking $f$ in re-ranking makes FLORA computationally similar to other ANN methods. However, we will empirically show that the performance of the discrete-space ranking (i.e., \textbf{FLORA}) closely resembles the performance of the ranking procedure that combines FLORA and a re-ranking step (denoted as \textbf{FLORA-R}).

\begin{figure*}	[h]
\vspace{0.05in}
\centering
\mbox{
    \includegraphics[width=2.8in]{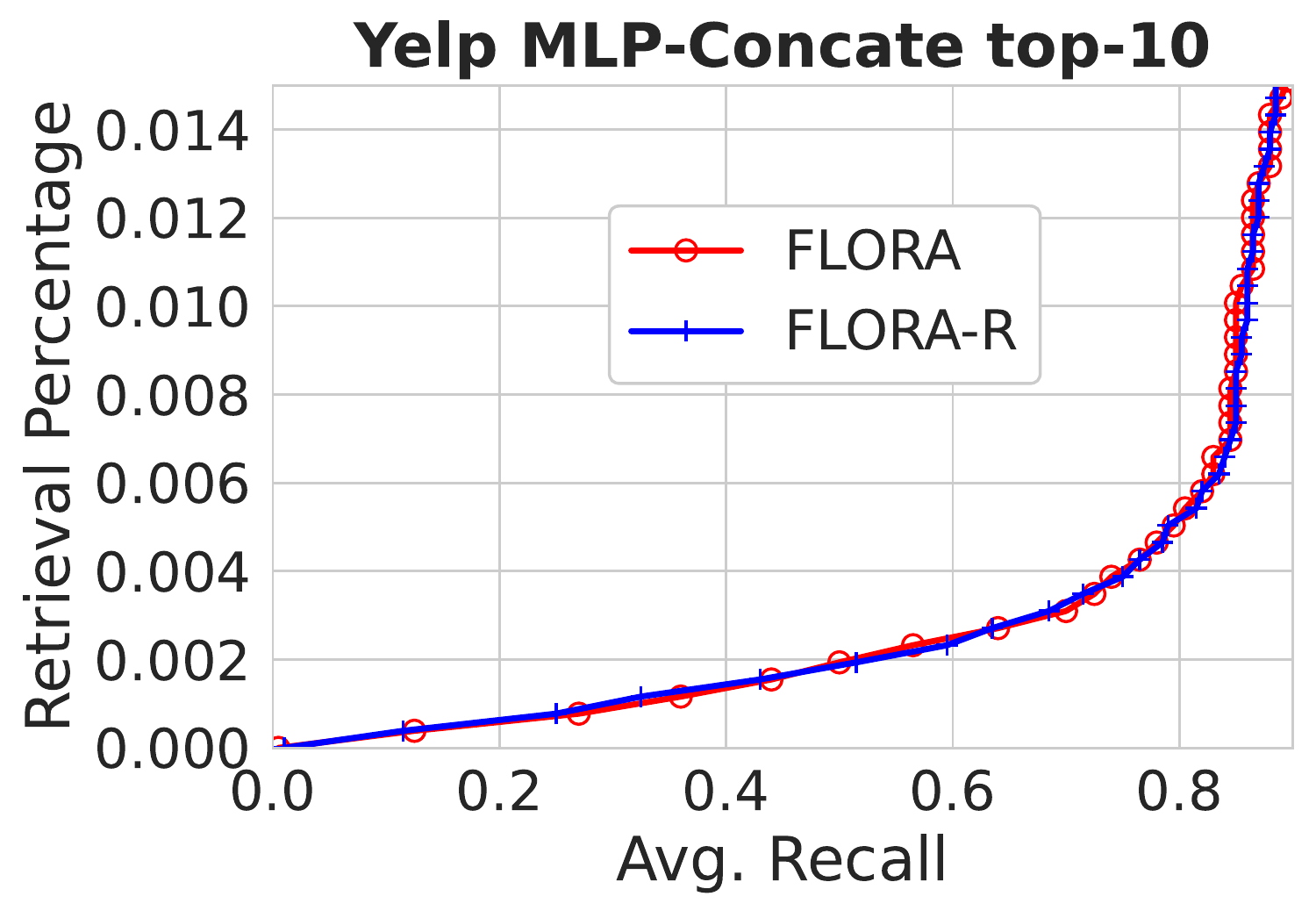}
    \hspace{0.2in}
	\includegraphics[width=2.8in]{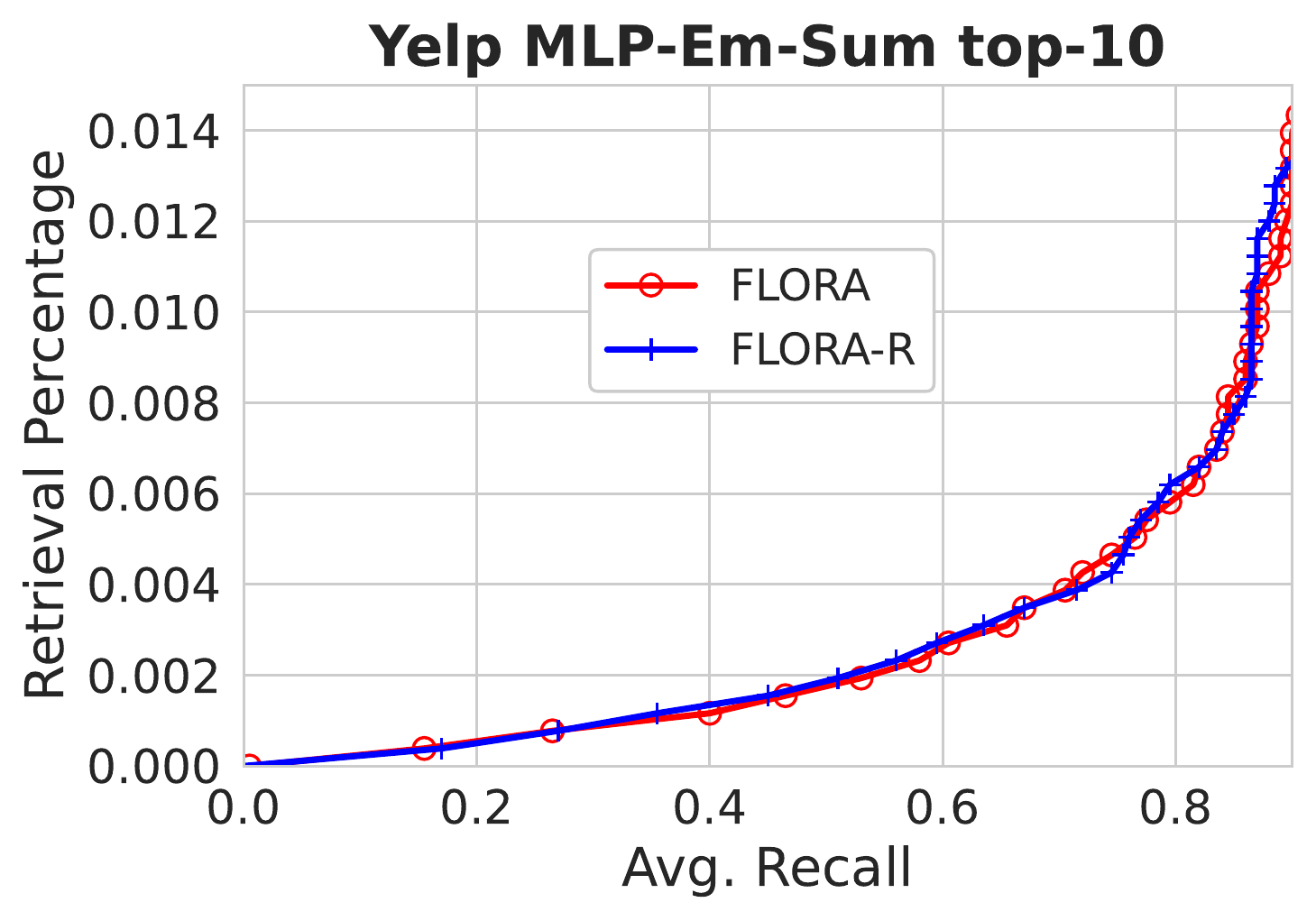}
}
\mbox{
	\includegraphics[width=2.8in]{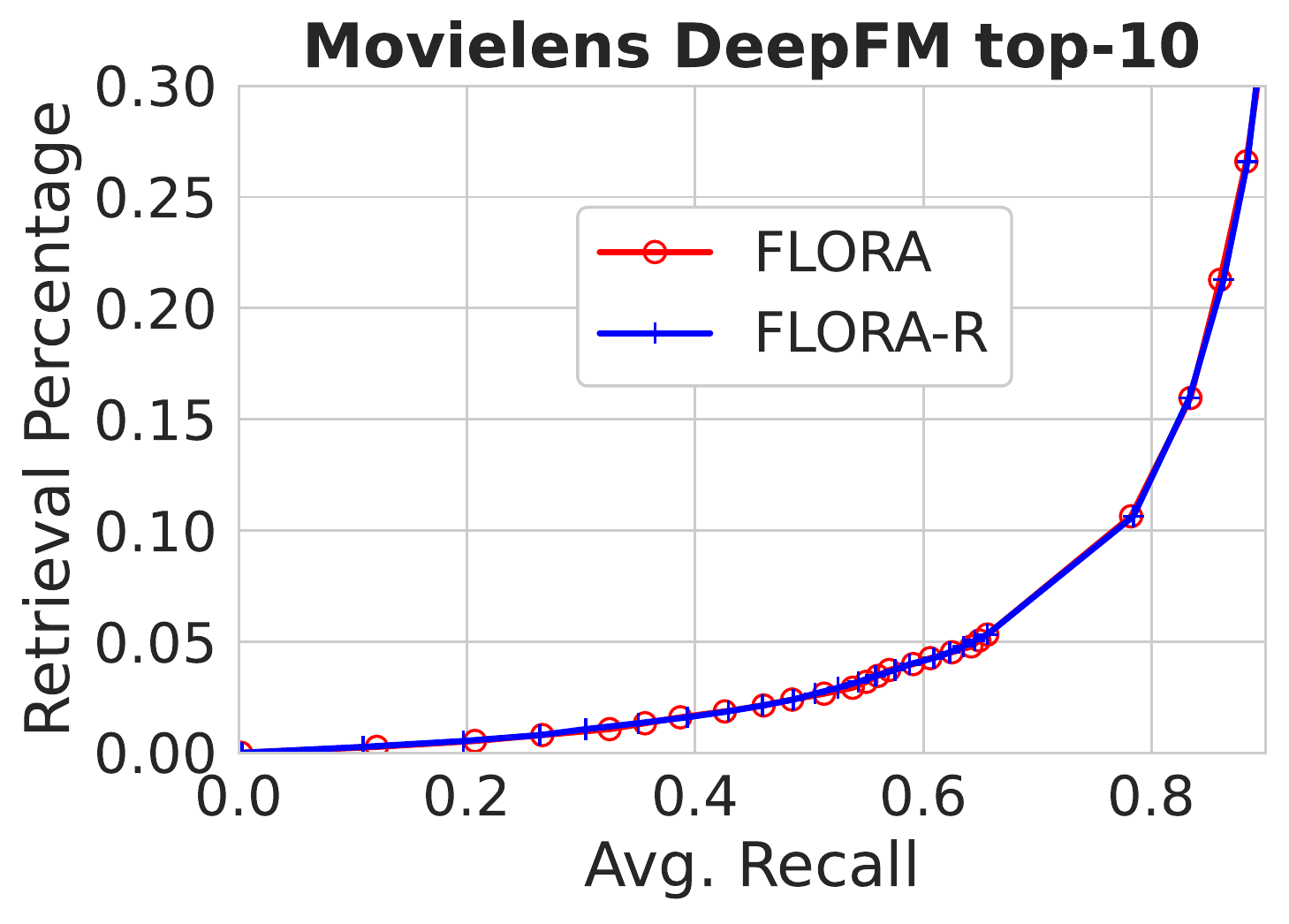}
    \hspace{0.2in}
	\includegraphics[width=2.8in]{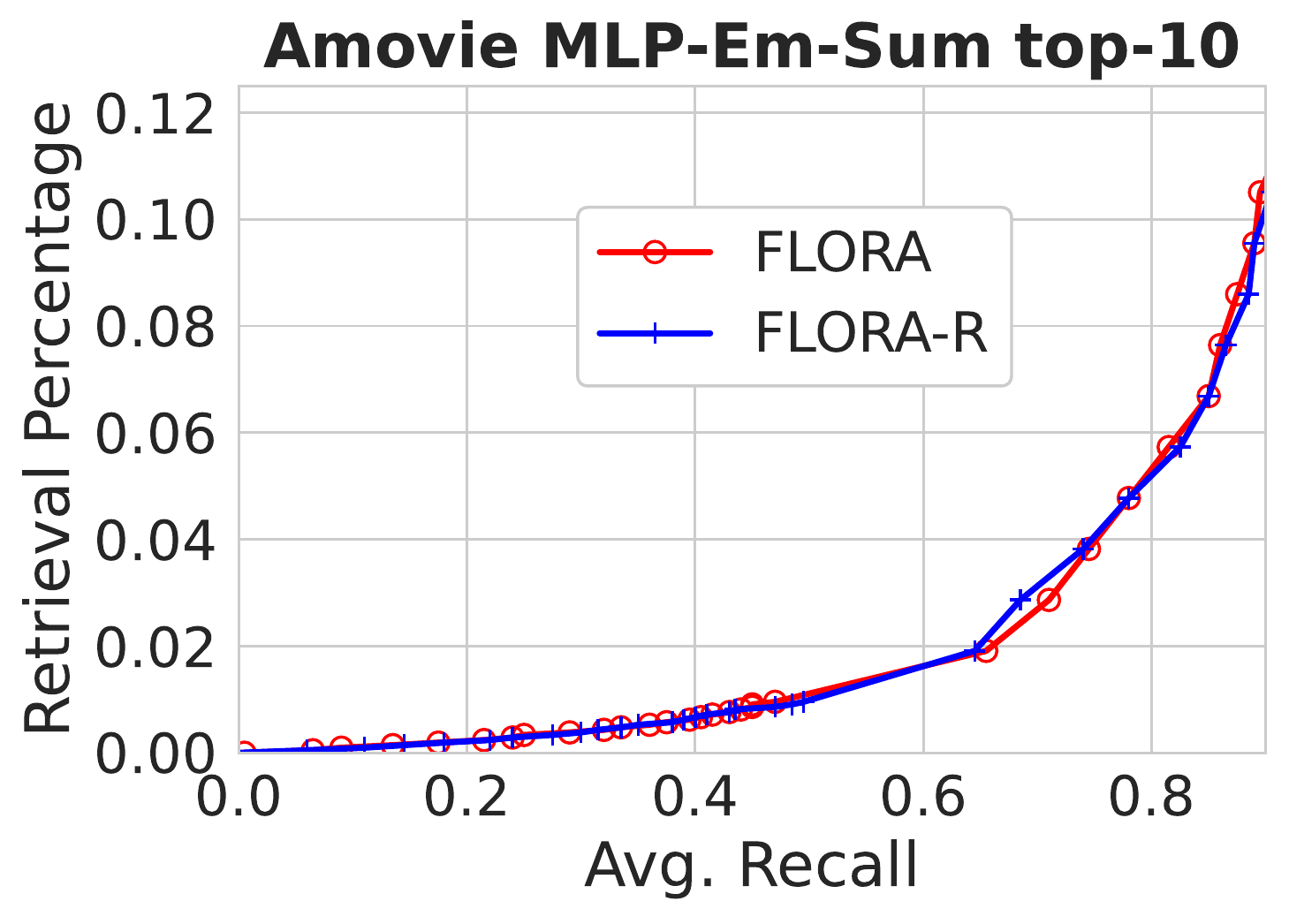}
}
    \vspace{-0.1in}
	\caption{Top-10 Recall results of  FLORA with and without the Re-ranking step.}\label{fig:results_recall_hash_reranking}\vspace{0.2in}
\end{figure*}

Figure~\ref{fig:results_recall_hash_reranking} shows the recall values at different retrieval percentages of the items for FLORA and FLORA-R. In FLORA-R, if there are ties at a specific threshold in the ranking list, the percentage is calculated using the size of the list that includes all the ties. Consequently, FLORA-R is expected to be always better than FLORA. As can be observed, there is not a noticeable difference between the two recall curves. This means that the discrete space ranking can be reliably used without access to the binary function $f$.

\newpage

\subsection{Improving Recall with More Hash~Tables}

\begin{figure*}[t!]
\centering
\mbox{
        \includegraphics[width=2.8in]{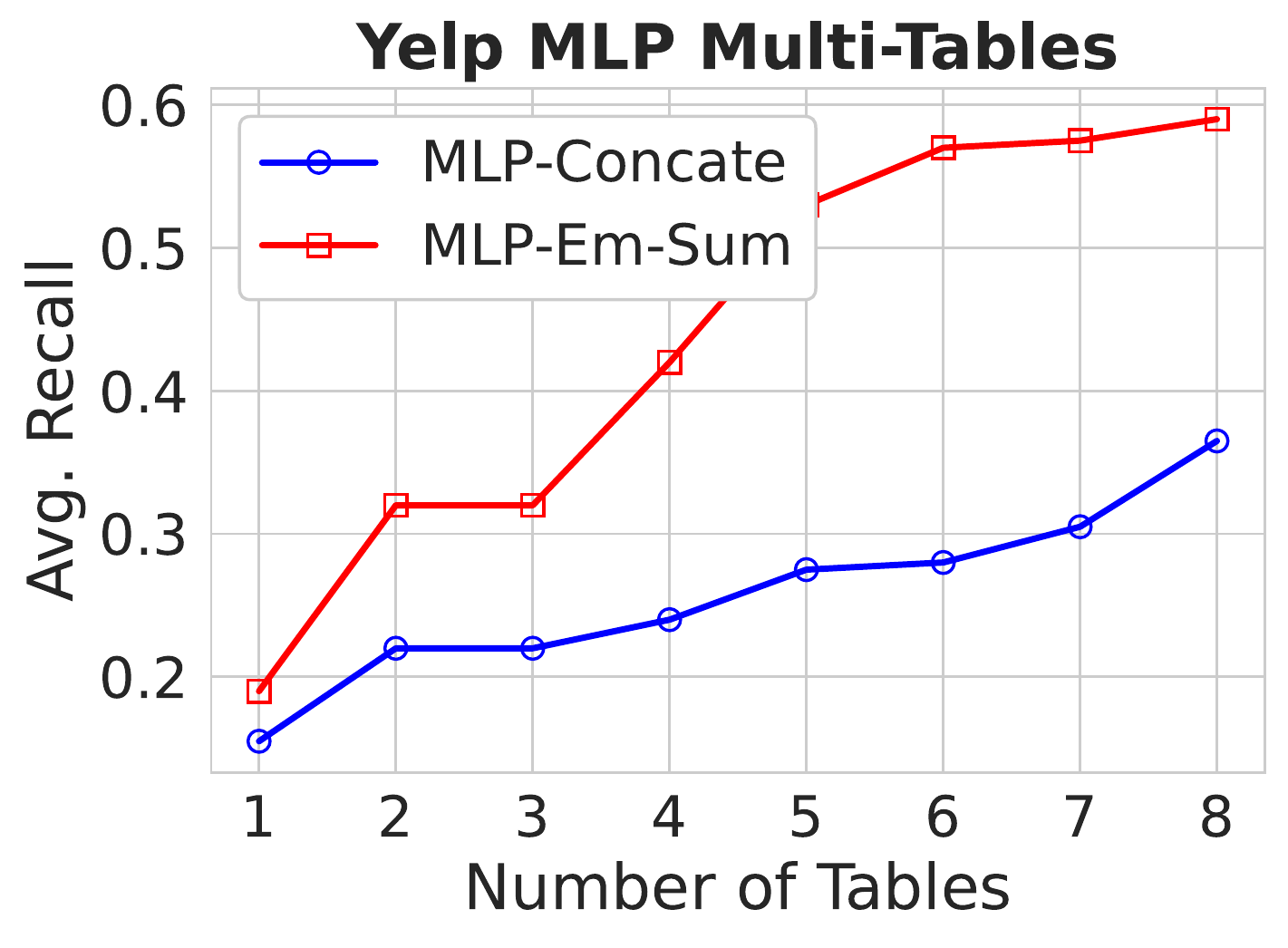}
        \hspace{0.2in}
        \includegraphics[width=2.8in]{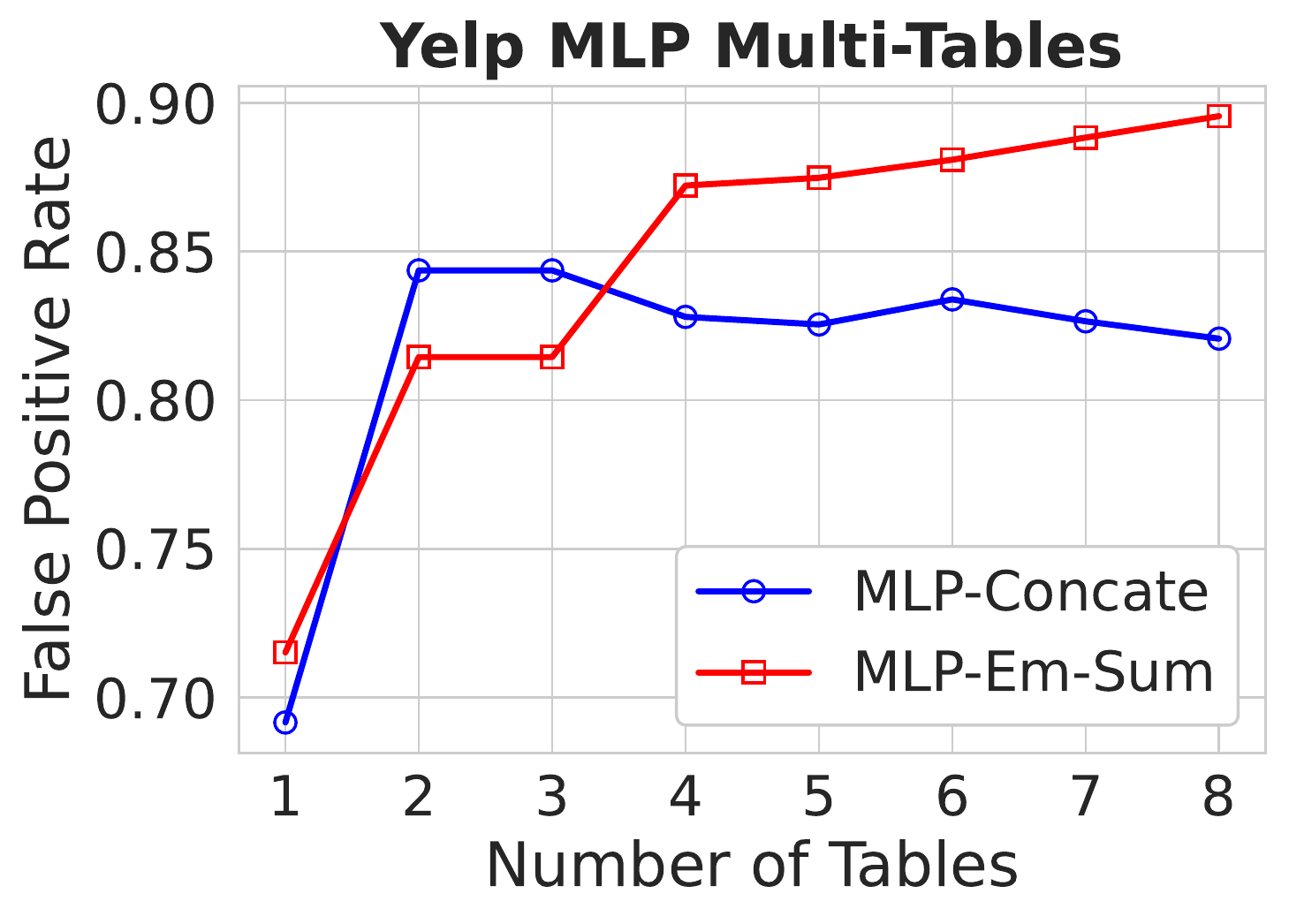}
}

\mbox{
		\includegraphics[width=2.8in]{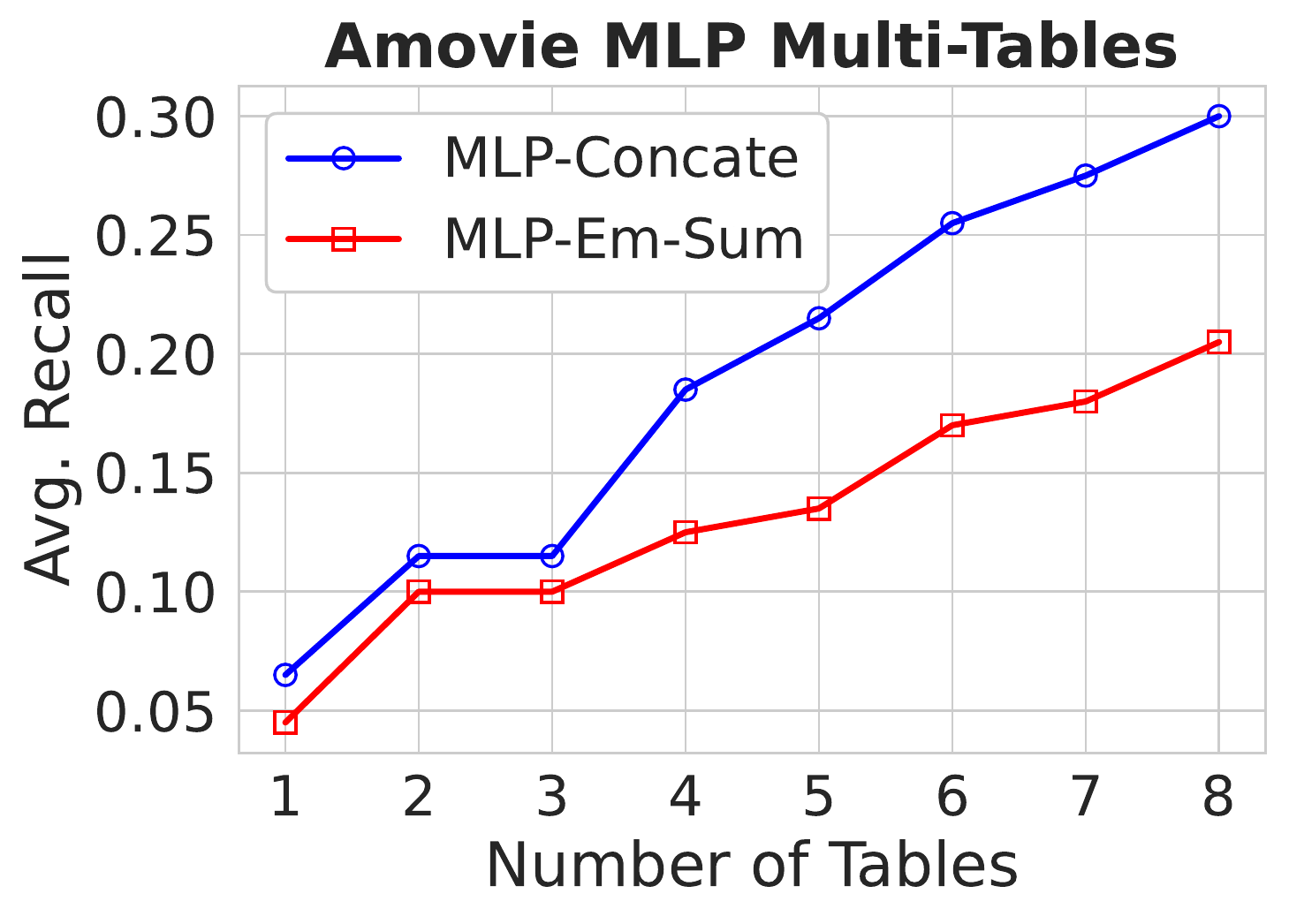}
        \hspace{0.2in}		
		\includegraphics[width=2.8in]{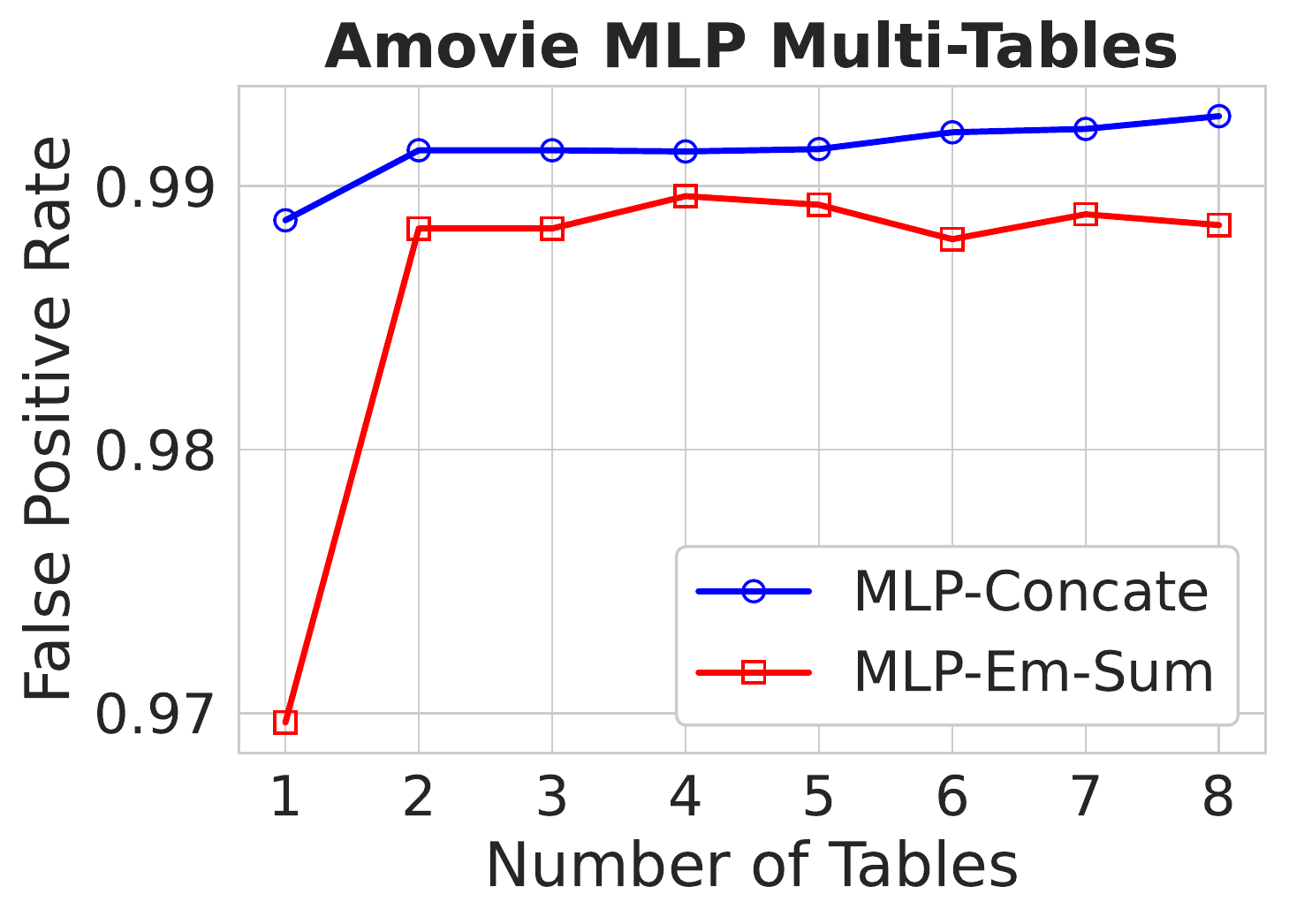}
}
    \vspace{-0.1in}
	\caption{FLORA with multiple hash tables: Top-10 recall and False Positive Rate of the candidates within radius-0 of the query. }\label{fig:results_retrieval_multitables}\vspace{-0.1in}
\end{figure*}

A principled approach to improve the probability of the relevant items in the candidate list is to employ multiple hash tables. Figure~\ref{fig:results_retrieval_multitables} shows the recall of the candidate items and their corresponding False Positive Rate (FPR) as multiple tables are used. We find candidate items whose hash codes are within distance 0 of the queries.

{As can be observed, the recall steadily increases as more hash tables are used. While FPR is expected to increase with more tables, empirically, we observe that it only increases slightly after 2 or more tables are used.} This feature of FLORA is an advantage over other methods. In practice, we can employ multiple hash tables to reach the desired recall rate.


\subsection{Effect of Sampling on Performance}\label{sec:result_sampling_efficiency}

\begin{figure}[t]
\centering
\mbox{
		\includegraphics[width=2.8in]{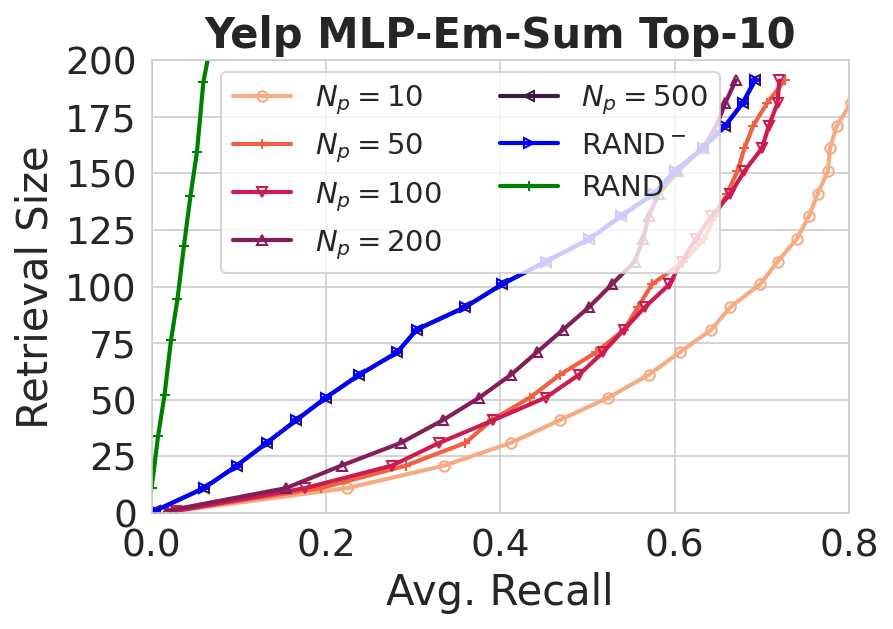}\hspace{0.2in}
		\includegraphics[width=2.8in]{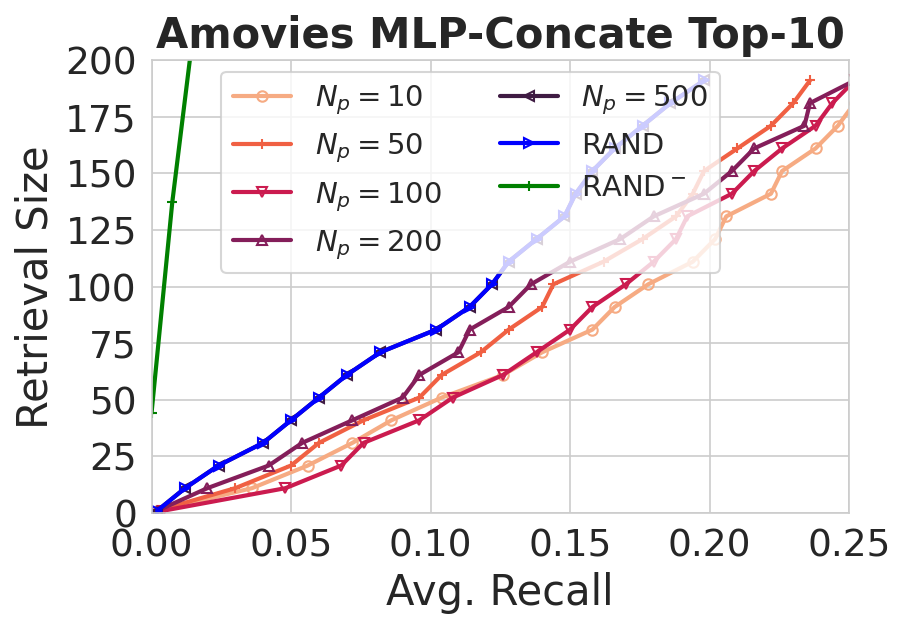}
}

\vspace{-0.1in}

	\caption{Effect of Sampling Strategies in FLORA}\label{fig:results_training_samplin}
     \vspace{0.3in}
\end{figure}

We discuss the importance of an effective sampling strategy in training. Figure~\ref{fig:results_training_samplin} shows the recall performance of different strategies. \textsc{Rand} refers to random selection of training pairs (Option 1 in Figure~\ref{fig:fltr_sampling}(a)). \textsc{Rand}$^-$ refers to Option 2 in Figure~\ref{fig:fltr_sampling}(b), where the positive list contains 10 items and the negative list has the remaining items; each list is randomly selected with probability 0.5, and items in the negative list are uniformly selected. Finally, the remaining experiments employ Option 3 in Figure~\ref{fig:fltr_sampling}(c), where $N_p$ denotes the size of the positive list; items in the negative list are selected with probability proportional to their similarity scores (with respect to the user) defined by binary function $f$. The recall of \textsc{Rand} is significantly worse than the other strategies. While \textsc{Rand}$^-$ improves the recall performance over \textsc{Rand}, it is more effective to adopt the proposed Option 3 (Figure~\ref{fig:fltr_sampling}(b)). Furthermore, as we shall show later, this strategy allows the training process to quickly reach the optimal~model.


\subsection{Ablation Study}\label{sec:result_ablation}
\begin{figure}[t!]
\centering
\mbox{
    \includegraphics[width=2.8in]{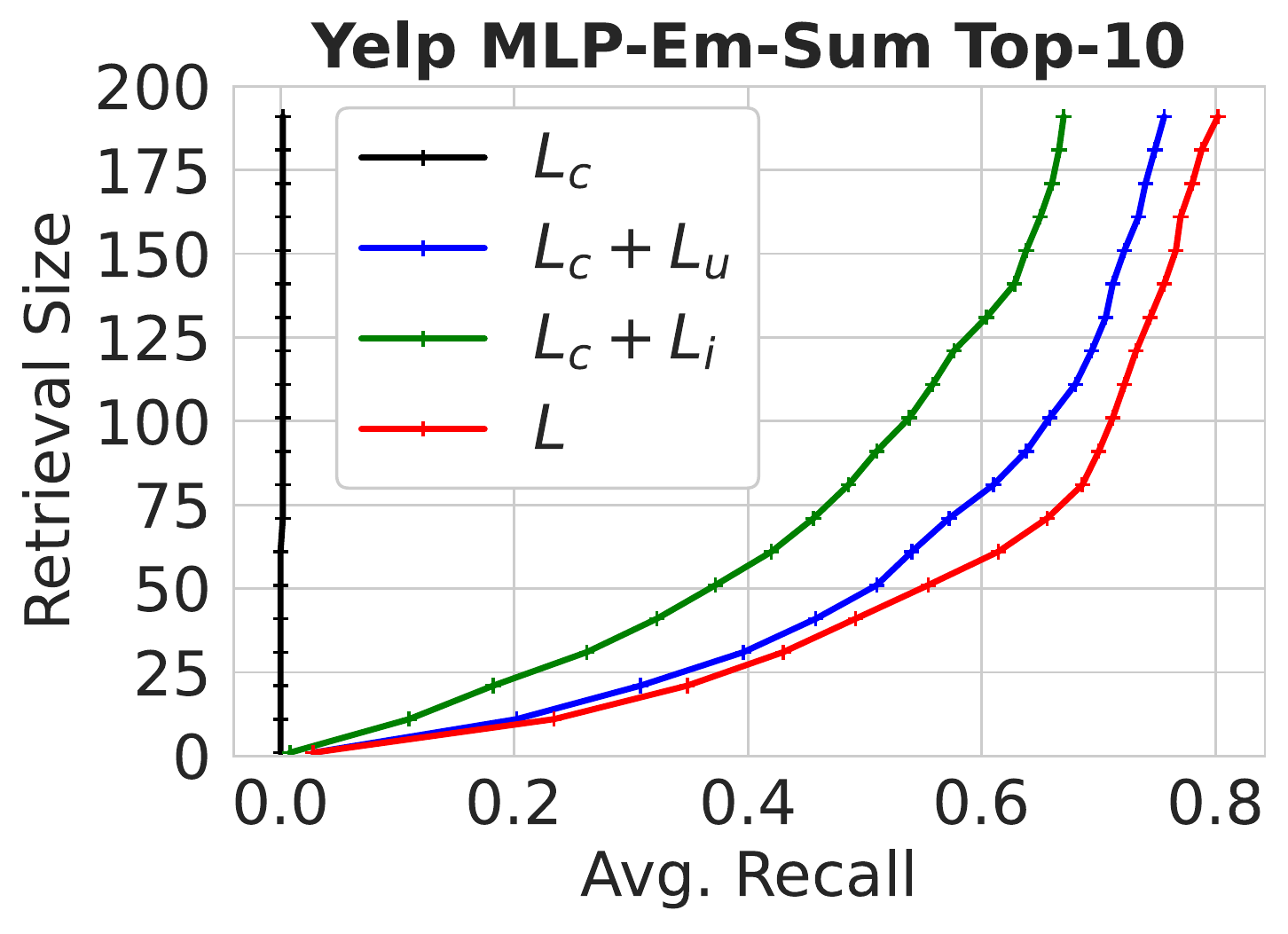}
    \hspace{0.2in}
    \includegraphics[width=2.8in]{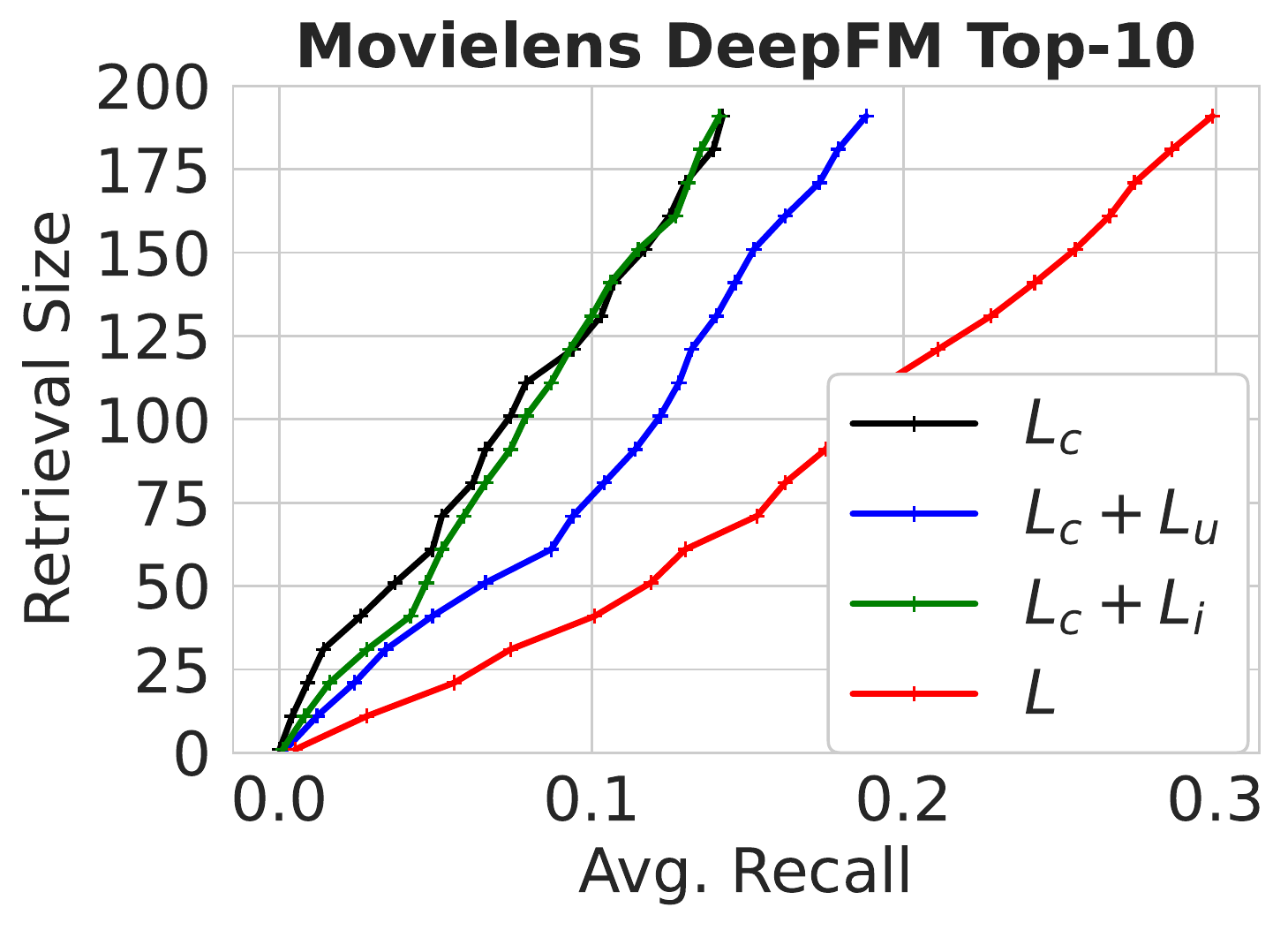}
}

\vspace{-0.1in}

	\caption{Recalls for different variants of the objectives.}\label{fig:results_ablation}
\end{figure}

We quantify the performance gains with respect to various objectives in FLORA through an ablation study. Figure~\ref{fig:results_ablation} shows the recall performance for different variants of the loss function, as proposed in Eq.~\eqref{eqn:overall_objective}. Without enforcing uniform frequency and independent bits, the recall performance drops significantly ($L_c$). Adding uniform frequency ($L_c+L_u$) results in a performance closer to that of the final model. While adding independent bits to $L_c$, resulting in the model $L_c+L_i$, does not result in a noticeable performance gain, the presence of $L_i$ in the final model (Eq.~\eqref{eqn:overall_objective}) improves the recall as the size of the retrieval list grows.

We conjecture that uniform frequency ensures that hash code learning does not fall into a collapsed-bit solution, where the training process fails to learn a specific bit, and thus it assigns most examples into the same bit value. While this is not necessarily serious, especially when the chosen size of the hash codes is large, this worsens the efficiency of the method because this means that similar performance can be achieved with smaller hash codes.

\newpage

\subsection{Training Efficiency}
\begin{figure}[ht!]
\centering
\mbox{
    \includegraphics[width=2.8in]{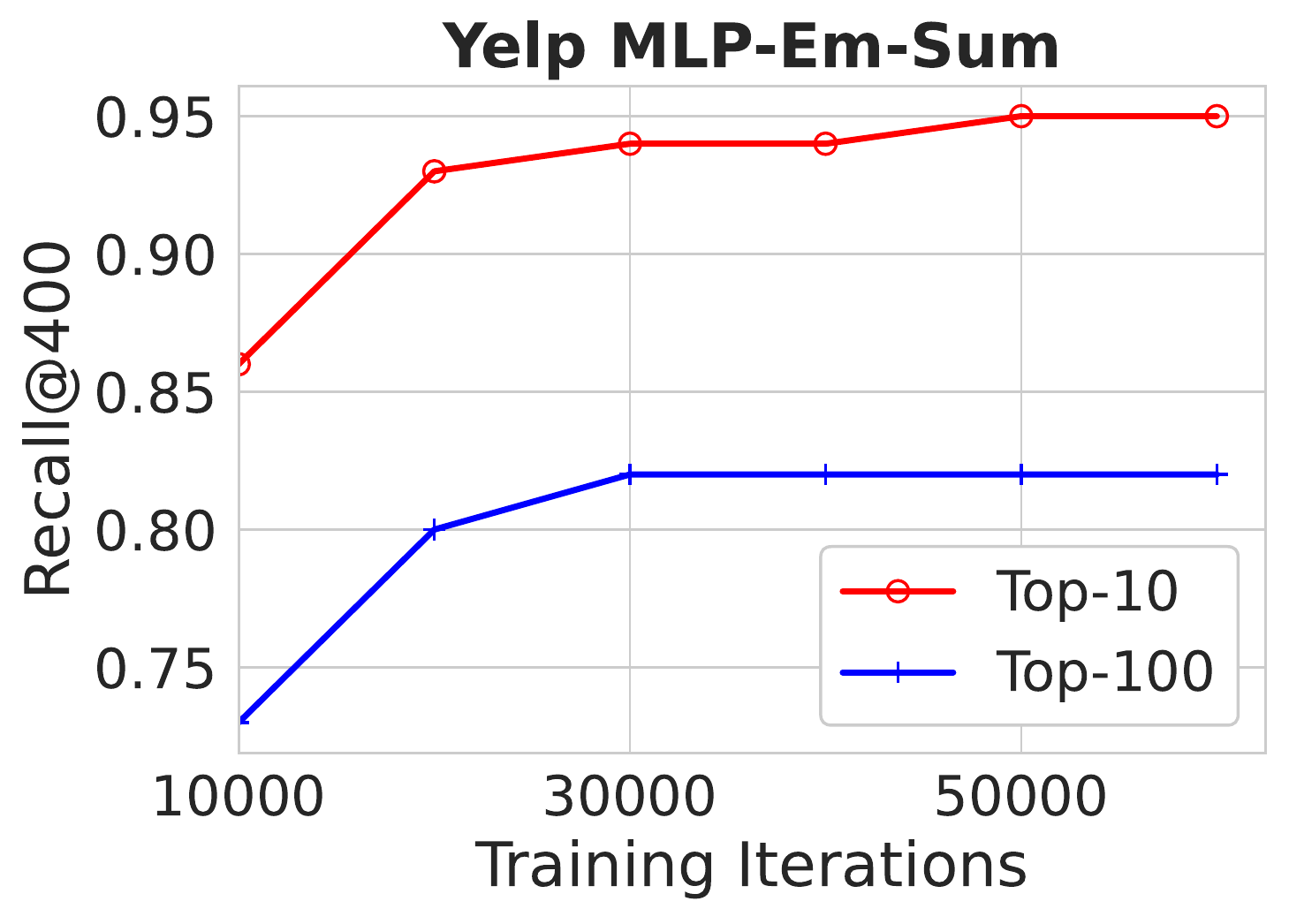}
    \hspace{0.2in}
    \includegraphics[width=2.8in]{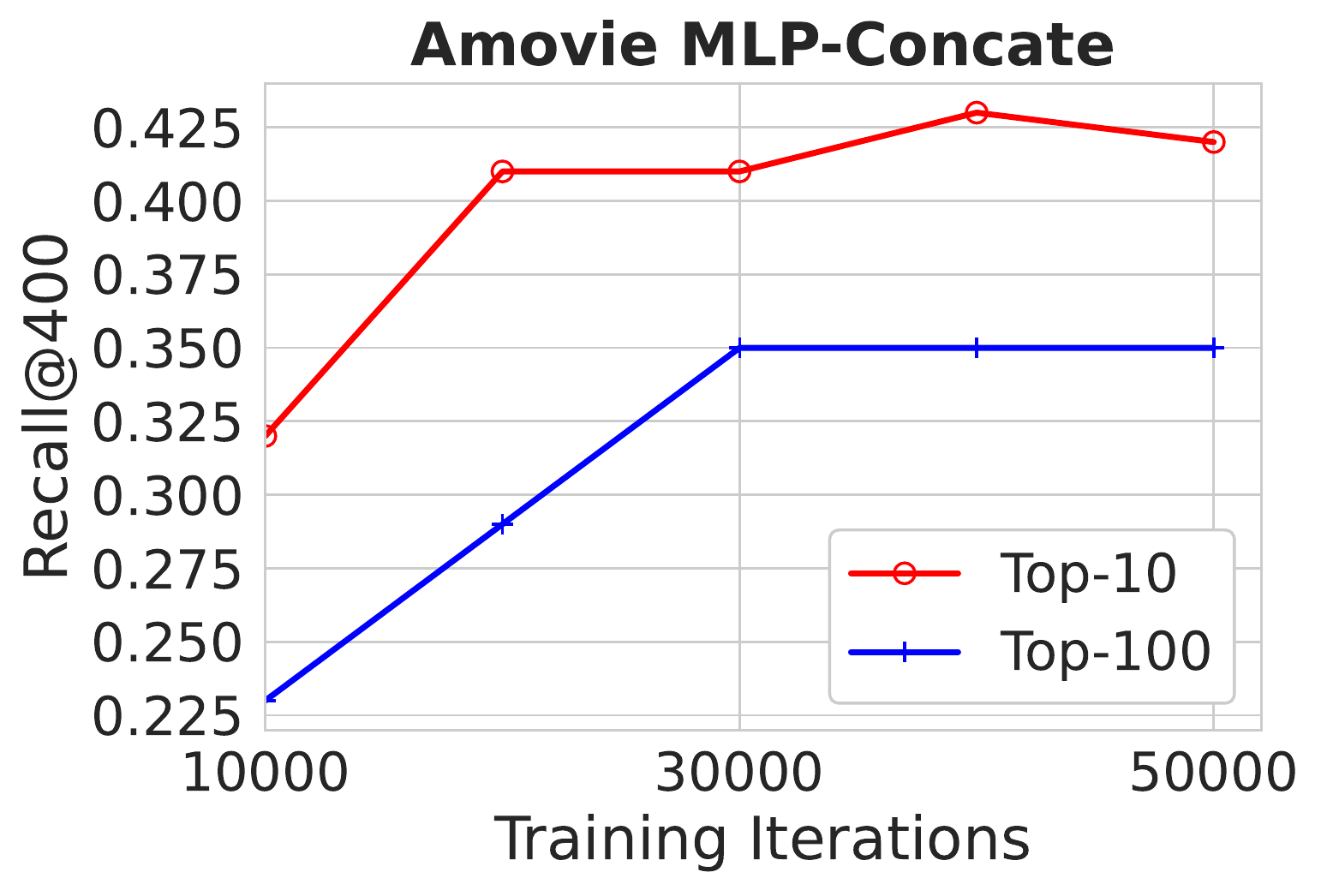}
}

	\caption{Recalls during model training. }\label{fig:results_training_efficiency}
\end{figure}
We present the recall performance during the model training in Figure~\ref{fig:results_training_efficiency}. As can be observed, after 20,000 iterations, the models have already approached the optimal performance. Compared to the results of \textsc{RAND} and \textsc{RAND}$^-$ in Figure~\ref{fig:results_training_samplin}, the proposed sampling strategy (Option 3 in Figure~\ref{fig:fltr_sampling}) is significantly more efficient in selecting relevant pairs for training. In practice, while there is a trade-off between training time and ranking performance in the other sampling strategies, even at a shorter training time, FLORA's proposed sampling strategy allows us to reach the optimal model in a significantly shorter amount of time.

\section{Conclusion}\label{sec:conclusion}

In this paper, we studied the fast neural ranking problem and proposed an approximate fast ranking framework, FLORA, based on an asymmetric hashing scheme. The proposed FLORA framework learns a pair of hash functions for two heterogeneous domains(e.g., users and items). The hashing model learns to preserve the original similarity of the heterogeneous objects through inner product fitting of the corresponding binary codes produced by the hash functions. Under this learning objective, the inner product, or equivalently the Hamming distance, of the learned binary codes can approximately reveal the true similarity defined by any binary function, including complex neural network functions.

Then, we propose an efficient sampling procedure to accelerate the training process and find better local optima of the proposed hashing model in FLORA. This procedure divides the items into two lists, positive and negative, where we first pick each list with equal probability, then an item in the negative list is selected with a probability inversely proportional to their rank defined by the neural network-based function, $f$.

Finally, we evaluate FLORA on several benchmark recommendation datasets and neural network functions  with different levels of complexity. FLORA is significantly better than existing approaches in both computation and ranking performance. We believe that FLORA can be used for other challenging problems which require a fast ranking phase, such as retrieval based question answering models. However, we leave this for future work.


\bibliographystyle{plainnat}
\bibliography{refs_scholar}

\end{document}